\documentclass[reprint,aps,prl,footinbib,floatfix,twocolumn,longbibliography,superscriptaddress,nobalancelastpage]{revtex4-2}
\usepackage{times}
\usepackage{amssymb}
\usepackage{color}
\usepackage{amsmath}
\usepackage{amsbsy}
\usepackage{amsthm}
\usepackage{dsfont}
\usepackage{graphicx}
\usepackage{bbm}
\usepackage{bm}
\usepackage{epsfig}
\usepackage{xfrac}
\usepackage{xcolor}
\usepackage{enumerate}
\usepackage{multirow}
\usepackage{physics}
\usepackage[T2A,T1]{fontenc}
\usepackage{ tipa }
\usepackage[english]{babel}
\usepackage{symbols}
\usepackage{makecell}
\usepackage{appendix}
\usepackage{dsfont}
\usepackage{float}
\usepackage{pdfpages}
\usepackage{caption}
\usepackage{subcaption}
\usepackage{ragged2e}
\usepackage{bbold}
\usepackage{placeins}
\usepackage{relsize}
\usepackage{xcolor}
\usepackage{scalerel}
\usepackage{tikz}
\usetikzlibrary{svg.path}
\definecolor{orcidlogocol}{HTML}{A6CE39}
\tikzset{
  orcidlogo/.pic={
    \fill[orcidlogocol] svg{M256,128c0,70.7-57.3,128-128,128C57.3,256,0,198.7,0,128C0,57.3,57.3,0,128,0C198.7,0,256,57.3,256,128z};
    \fill[white] svg{M86.3,186.2H70.9V79.1h15.4v48.4V186.2z}
                 svg{M108.9,79.1h41.6c39.6,0,57,28.3,57,53.6c0,27.5-21.5,53.6-56.8,53.6h-41.8V79.1z M124.3,172.4h24.5c34.9,0,42.9-26.5,42.9-39.7c0-21.5-13.7-39.7-43.7-39.7h-23.7V172.4z}
                 svg{M88.7,56.8c0,5.5-4.5,10.1-10.1,10.1c-5.6,0-10.1-4.6-10.1-10.1c0-5.6,4.5-10.1,10.1-10.1C84.2,46.7,88.7,51.3,88.7,56.8z};
  }
}
\newcommand\orcid[1]{\href{https://orcid.org/#1}{$\,$\mbox{\scalerel*{
\begin{tikzpicture}[yscale=-1,transform shape]
\pic{orcidlogo};
\end{tikzpicture}
}{|}}}}
\usepackage{soul}

\definecolor{myurlcolor}{rgb}{1,0,1}
\definecolor{myrefcolor}{rgb}{1,0,1}
\usepackage[colorlinks]{hyperref}
\hypersetup{unicode=true,
    bookmarksopen=false,
    breaklinks=false,
    pdfborder={0 0 0},
	bookmarksnumbered=false,
	pdfstartview={FitH},
	citecolor={myurlcolor},
	linkcolor={myrefcolor},
	urlcolor={myurlcolor}}
\makeatletter
\AtBeginDocument{\let\LS@rot\@undefined}
\makeatother
\makeatletter
\def\maketitle{
\@author@finish
\title@column\titleblock@produce
\suppressfloats[t]}
\makeatother
\renewcommand\bra[1]{{\langle{#1}|}}
\renewcommand\ket[1]{{|{#1}\rangle}}
\newtoggle{arXiv}
\toggletrue{arXiv}       % you DO want to see it
\begin{document}
\title{Spectrally enhanced gravitational redshift signals in photonic interferometry
%Probing Superposed Gravitational Redshifts with Cavity-QED-Enhanced Interferometry
}
\author{Jerzy Paczos\orcid{0000-0002-0674-9819}}
\email{jerzy.paczos@fysik.su.se}
\affiliation{Department of Physics, \href{https://ror.org/05f0yaq80 }{Stockholm University}, SE-106 91 Stockholm, Sweden}

\author{Piotr T. Grochowski\orcid{0000-0002-9654-4824}}
\email{piotr.grochowski@upol.cz}
\affiliation{Department of Optics, \href{https://ror.org/04qxnmv42}{Palacký University}, 17. listopadu 1192/12, 771 46 Olomouc, Czech Republic}

\author{Natalia Bruno\orcid{0000-0002-5694-4866}}
\email{bruno@lens.unifi.it}
\affiliation{\href{https://ror.org/02dp3a879}{CNR - Istituto Nazionale di Ottica (CNR-INO)}, Firenze, Italy}
\affiliation{\href{https://ror.org/04x48z588}{LENS - European Laboratory for Non-Linear Spectroscopy}, Sesto Fiorentino (FI), Italy}

\author{Emanuele Distante\orcid{0000-0001-5728-759X}}
\email{emanuele.distante@unifi.it}
\affiliation{Dipartimento di Fisica e Astronomia, \href{https://ror.org/04jr1s763}{Università degli Studi di Firenze}, Via Sansone 1, 50019 Sesto Fiorentino (Firenze), Italy}
\affiliation{\href{https://ror.org/04x48z588}{LENS - European Laboratory for Non-Linear Spectroscopy}, Sesto Fiorentino (FI), Italy}
\author{Stephan Welte\orcid{0009-0000-3743-7394}}
\email{stephan.welte@pi5.uni-stuttgart.de}
\affiliation{\href{https://ror.org/01z25am55 }{5. Physikalisches Institut and Center for Integrated Quantum Science and Technology}, \href{https://ror.org/04vnq7t77}{Universität Stuttgart}, Pfaffenwaldring 57, 70569 Stuttgart, Germany}

\author{Magdalena Zych\orcid{0000-0002-8356-7613}}
\email{magdalena.zych@fysik.su.se}
\affiliation{Department of Physics, \href{https://ror.org/05f0yaq80 }{Stockholm University}, SE-106 91 Stockholm, Sweden}

\begin{abstract}
    \noindent 
    Quantum mechanics and general relativity are remarkably successful, yet their joint predictions remain beyond direct experimental reach.
    Photonic interferometers operating in a gravitational field offer a promising route to test this regime: gravitational redshift differentiates photon propagation between vertically separated arms, leading to a predicted phase shift and reduction in interference visibility.
    These effects are minute on laboratory scales, motivating existing proposals to consider large interferometer areas to accumulate them geometrically.
    Indeed, such schemes typically require kilometre-scale optical paths to produce appreciable interferometric signals.
    Here, we propose a strategy that replaces geometric accumulation with spectral enhancement by coupling the two interferometer arms to optical scatterers based on ultranarrow atomic transitions.
    These transitions serve as local spectral references: gravitational redshift changes the photon's detuning from each resonance, producing large differences in reflection amplitude and hence substantial visibility loss and interferometric phase shift already for centimetre-scale height differences.
    We quantify the experimental requirements and assess the feasibility of implementation with current or near-term technology.
    Our proposal thus opens a route to tabletop optical tests of the interplay between quantum mechanics and general relativity.
\end{abstract}
\maketitle
\twocolumngrid

\noindent The interplay between quantum mechanics and general relativity remains one of the most intriguing yet least experimentally explored frontiers of modern physics.
In general relativity, elapsed time is not a universal parameter but a local quantity accumulated along a worldline, giving rise to gravitational time dilation and redshift effects~\cite{Einstein1916} that have been verified in numerous experiments~\cite{Pound1960, Hafele1972, Chou2010, McGrew2018, Bothwell2022, Zheng2022}.
Combined with quantum mechanics, this opens up a distinct regime in which a single system can evolve in a coherent superposition of paths associated with different worldlines, and therefore with different accumulated proper times~\cite{Sinha2011, Zych2011}.
This regime becomes accessible in single-particle quantum interferometry, where proper-time differences lead to phase shifts, path distinguishability, and loss of interference visibility~\cite{Sinha2011, Zych2011, Pikovski2015}.
Single-photon interferometers provide a clean setting to access this regime, as gravitational time dilation directly affects the optical phase and the temporal overlap of optical modes, thereby affecting interference visibility~\cite{Zych2012, Bruschi2023}.
Recent progress in large-scale single-photon interferometry suggests that photonic tests of relativistic quantum interference are becoming a concrete experimental direction~\cite{Wu2024, Yu2026}.

\begin{figure}[h!t]
    \centering
    \includegraphics[width=1.0\linewidth]{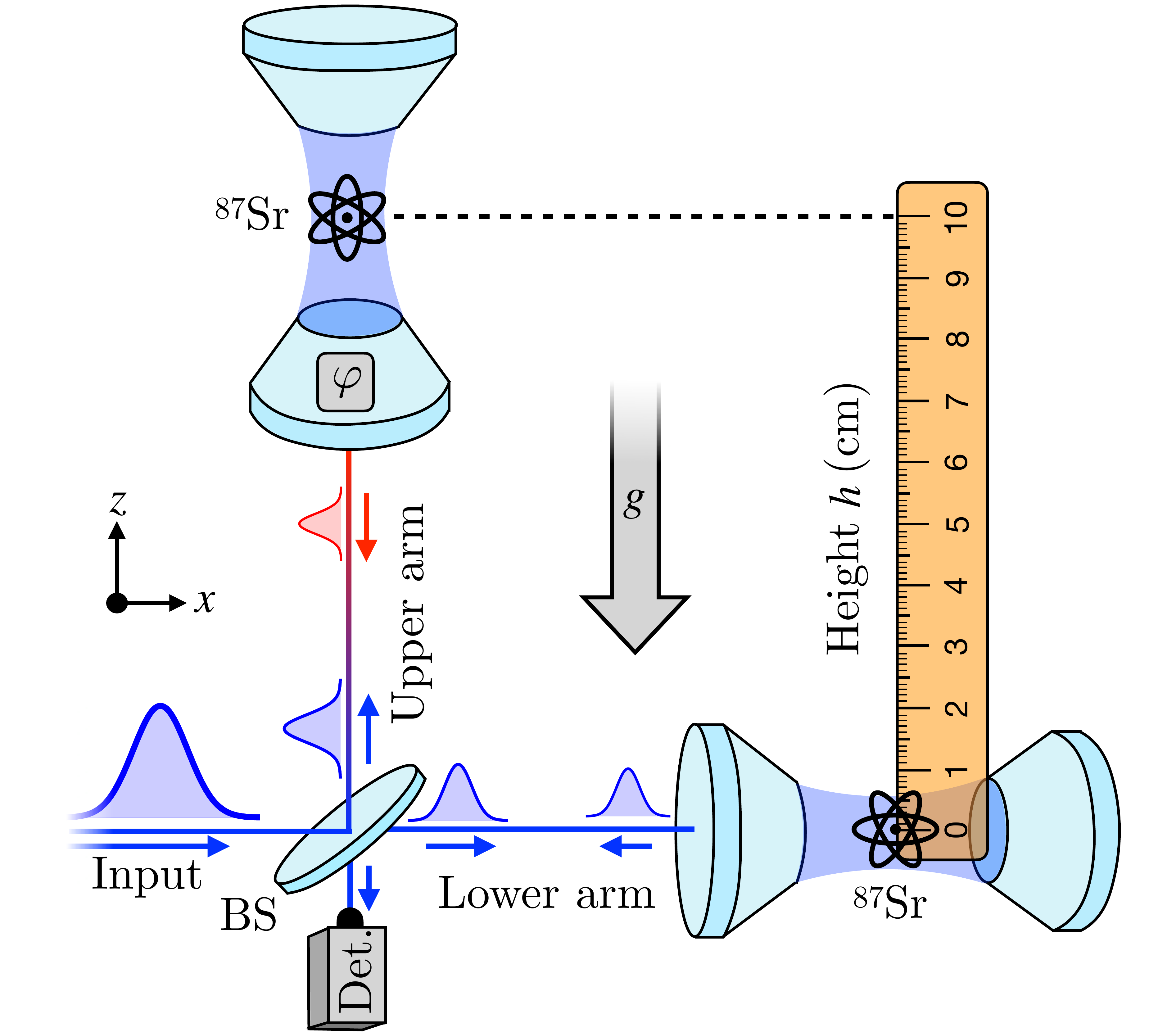}
    \caption{\justifying
        \textbf{Cavity-quantum-electrodynamics-enhanced photonic interferometry in a gravitational field.} An input photon is split at a beam splitter (BS) between the arms of a Michelson interferometer, which terminate in identical optical scatterers positioned at different heights, each implemented as a cavity coupled to an ultranarrow atomic transition. Due to the gravitational redshift, the two otherwise identical scatterers are probed at different detunings; the reflected amplitudes therefore differ in magnitude and phase, producing a gravitationally induced change in the output interference visibility and phase. An externally controlled phase \(\varphi\) in the upper arm is used to scan the fringes. 
}
\label{fig: Michelson}
\end{figure}

First photonic proposals and current experimental efforts, however, rely on geometric amplification of the small gravitational-redshift effects~\cite{Zych2012, Hilweg2017, Mohageg2022, Mohageg2025, Yu2026, Wu2024}.
In such schemes, the gravitational effect accumulates during free propagation, so that the resulting phase difference scales as~\cite{Zych2012}
\begin{equation}\label{free propagation phase}
    \Phi_g\sim\omega_0\frac{l}{c}\frac{gh}{c^2},
\end{equation}
where $\omega_0$ is the optical frequency, c is the speed of light, $l$ is the propagation distance ($l/c$ is the propagation time), $h$ is the vertical separation between the interferometer arms, and $g$ is the gravitational acceleration.
Since laboratory-scale height differences give an extremely small fractional gravitational redshift $gh/c^2$, appreciable signals require extremely long propagation arms~\cite{Zych2012}.
This pushes photonic implementations towards kilometre-scale optical paths, where path-length noise, dispersion, thermal fluctuations, and phase drift become dominant experimental challenges~\cite{Yu2026, Wu2024}.
Related photonic proposals based on photon-mediated entanglement, quantum memories, and Hong--Ou--Mandel interferometry reduce the required propagation length by storing photonic excitations, obtaining the signal through proper-time accumulation over vertically separated memories with proposed near-term regimes involving metre- to hundred-metre height differences~\cite{Barzel2024, Borregaard2025, Gundogan2026}.

Here, we propose an alternative route that greatly reduces the required size of the setup, allowing implementation in a tabletop setting.
Our proposal augments a photonic interferometer with cavity-quantum-electrodynamics (cavity-QED) scatterers~\cite{Reiserer2015} based on ultranarrow clock transitions~\cite{Ludlow2015}, placed in its two arms (see Fig.~\ref{fig: Michelson}).
Because gravity shifts the photon frequency relative to each scatterer's resonance, otherwise identical scatterers are probed at slightly different detunings.
Near an ultranarrow resonance, this minute frequency shift is converted into a large differential optical response, producing path-dependent reflection or transmission amplitudes, and hence a gravitationally induced phase shift or path distinguishability at the output.
For optical cavities coupled to ultranarrow atomic transitions, the relevant scale is set by the linewidth of the atomic resonance.
When the gravitational frequency shift is small compared to the resonance linewidth, the resulting phase shift scales as
\begin{equation}\label{phase with scatterers}
    \Phi_g \sim Q\frac{gh}{c^2},
\end{equation}
where $Q=\omega_0/\gamma$ is the quality factor of the atomic transition and $\gamma$ is the associated polarization decay rate.
Equating this scaling with the free-propagation result, Eq.~\eqref{free propagation phase}, shows that achieving the same enhancement geometrically would require a propagation time $l/c\sim1/\gamma$.
For an ultranarrow linewidth $\gamma\sim1\,\mathrm{mHz}$, typical of optical-clock transitions~\cite{Ludlow2015}, this corresponds to an arm length $l\sim10^{11}\mathrm{m}$, comparable to the Earth--Sun distance.
This limit illustrates the origin of the enhancement: the small fractional redshift is converted into an amplified optical response not by increasing the propagation time but by resolving it spectrally.
On the other hand, centimetre-scale height differences can already make the gravitational frequency shift comparable to the resonance linewidth of ultranarrow clock transitions.
The interferometer then enters a regime in which the photon probes substantially different parts of the scattering spectrum in the two arms.
In this regime, suitable cavity-QED scatterers can produce large scattering asymmetries, resulting in complete loss of interference visibility, or gravity-induced phase shifts of order $\pi$ over only a few to tens of centimetres.
%In this way, our proposal replaces large-area free-propagation interferometry with a compact (order-10-cm) interferometric setup with a redshift-to-scattering transducer in each arm, implemented using a cavity coupled to a narrow atomic transition.

As in the experiment proposed in Ref.~\cite{Zych2012}, the setup proposed here aims to test gravitational effects on quantum interference.
When implemented with nonclassical states of light, the height-dependent gravitational modulation of single-photon interference will be incompatible both with classical electromagnetism on a curved background and with quantum optics subject to non-relativistic (Newtonian) gravity.
The experiment would therefore access the regime whose description requires both a quantum theory of light and the general-relativistic gravitational redshift, providing a compact test of their interplay.

\section*{Cavity-QED-enhanced interferometer}
We consider a single-photon Michelson interferometer in a uniform gravitational field, with its two arms terminated by identical optical scatterers positioned at heights differing by \(h\) (Fig.~\ref{fig: Michelson}).
At this stage, each scatterer is described phenomenologically by a complex reflection amplitude $r(\omega)$, where $\omega$ is the photon frequency measured locally at the scatterer.
The corresponding reflection probability and phase are $R(\omega)=|r(\omega)|^2$ and $\varphi(\omega)=\arg r(\omega)$.
Although the scatterers are identical, gravitational redshift makes the photon frequency seen by each scatterer slightly different.
If the photon frequency in the lower arm is $\omega_{\rm l}$, then the scatterer in the upper arm sees a redshifted frequency given by $\omega_{\rm u}=\omega_{\rm l}+\delta\omega_g$, where $\delta\omega_g\simeq -\omega_{\rm l}\frac{gh}{c^2}$, for $gh/c^2\ll1$.
Thus, gravitational redshift causes the two branches of the photon’s path superposition to probe different points of the same frequency-dependent scattering response.
As a result, the optical fields in the two arms after reflection also acquire different amplitudes
\begin{equation}
    r_{\rm l}=r(\omega_{\rm l}),\qquad r_{\rm u}=r(\omega_{\rm u}).
\end{equation}
The interference visibility and the gravity-induced phase shift are therefore (see Methods)
\begin{equation}\label{visibility and phase}
    V=\frac{2\sqrt{R_{\rm l}R_{\rm u}}}{R_{\rm l}+R_{\rm u}},
    \qquad
    \Phi_g=\phi_{\rm u}-\phi_{\rm l},
\end{equation}
where $R_{\rm l,u}=|r_{\rm l,u}|^2$ and $\phi_{\rm l,u}=\arg(r_{\rm l,u})$.
These expressions show that a gravitational redshift affects the interferometric signal whenever it changes the magnitude or the phase of the scattering amplitude across the two arms.
The magnitude of the effect is controlled by the spectral scale over which $r(\omega)$ varies, with sharper frequency dependence producing larger changes in visibility or phase for the same redshift.
In our proposed settings, an ultranarrow atomic transition coupled to an optical cavity provides exactly such a sharp spectral response.

\begin{figure}[h!t]
\includegraphics[width=\linewidth]{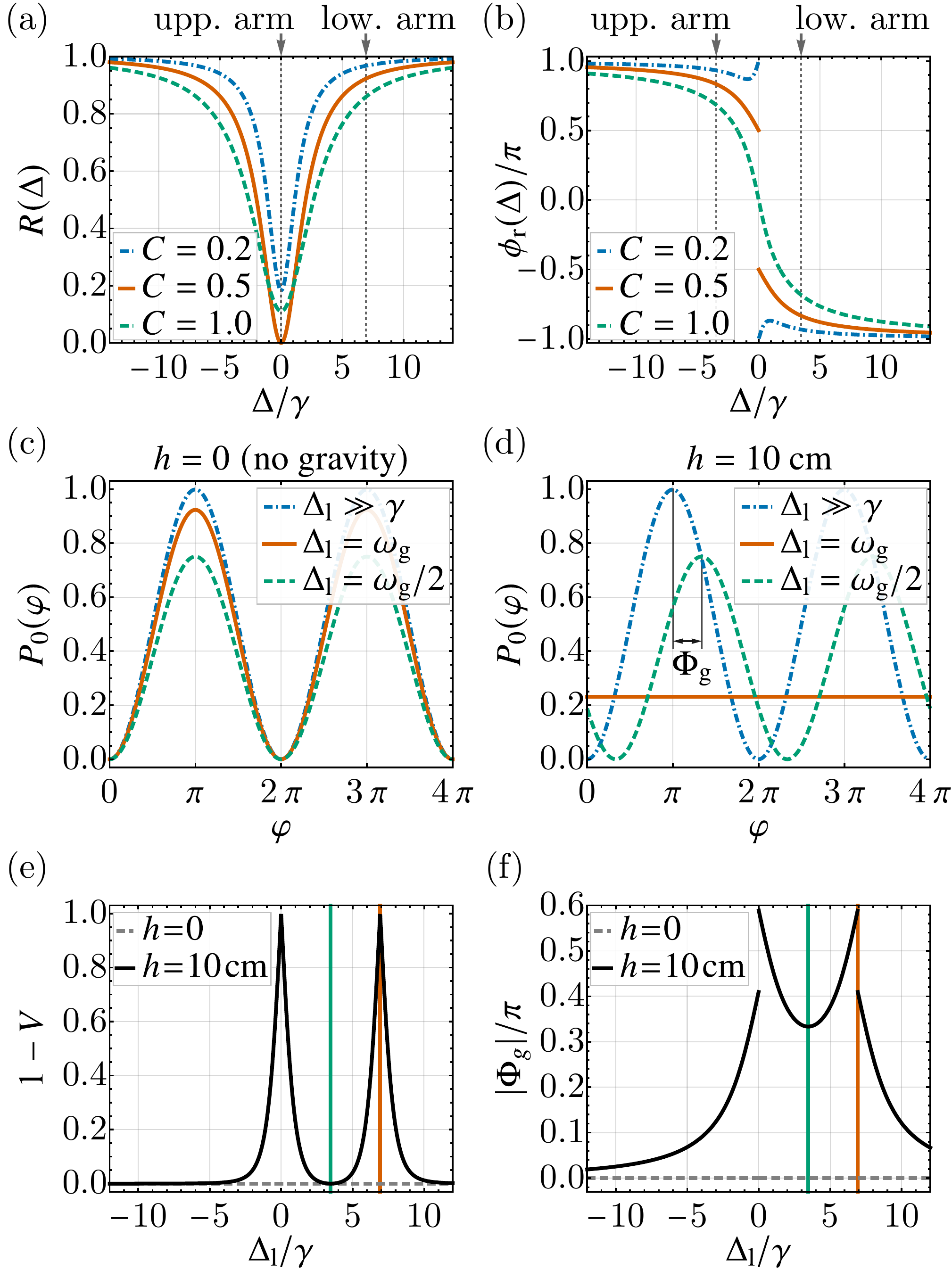}
\caption{\justifying
\textbf{Gravitational signatures in photon interference.}
(a) Reflection probability \(R(\Delta)\) and (b) reflection phase \(\phi_r(\Delta)\) of a single-sided cavity-QED scatterer versus detuning \(\Delta/\gamma\), for  \(C=0.2\), \(0.5\), and \(1\).
The dashed vertical lines represent the photon detuning in each arm for: (a) upper-arm resonance and (b) symmetric detuning.
(c),(d) Detection probability \(P_0(\varphi)\) at one output port as a function of the scanned interferometer phase \(\varphi\), for three lower-arm probe detunings: a far-detuned reference \(\Delta_{\rm l}\gg\gamma\) (blue), upper-arm-resonance \(\Delta_{\rm l}=\omega_g\) (orange), and symmetric detuning \(\Delta_{\rm l}=\omega_g/2\) (green), where \(\omega_g\equiv|\delta\omega_g|\) is evaluated at \(h=10\,{\rm cm}\).
(c) At \(h=0\), all settings yield full-contrast fringes with a common minimum.
(d) At \(h=10\,\)cm the reference is unchanged, the \(\Delta_{\rm l}=\omega_g\) fringe becomes flat (\(V\to0\), complete which-path information), and the \(\Delta_{\rm l}=\omega_g/2\) fringe retains full contrast but is shifted by \(\Phi_g\) (arrow).
(e) Visibility loss \(1-V\) and (f) phase \(|\Phi_g|/\pi\) as functions of \(\Delta_{\rm l}\) at \(h=0\) (grey dashed) and \(h=10\,\)cm (black); vertical lines mark the two working points \(\Delta_{\rm l}=\omega_g\) (orange) and \(\Delta_{\rm l}=\omega_g/2\) (green).
Panels (c)--(f) use the \({}^{87}{\rm Sr}\) clock transition and \(C=0.5\), giving \(\delta\omega_g/\gamma\simeq-6.93\) at \(h=10\,\)cm.}
\label{fig:gedanken}
\end{figure}

We work in the weak-excitation regime, where the atomic saturation can be neglected~\cite{Reiserer2015}, and restrict attention to the narrow spectral window around the common atom--cavity resonance.
Denoting this resonance frequency by $\omega_0$, we write $\Delta=\omega-\omega_0$ for the photon detuning, $g_0$ for the single-photon atom--cavity coupling rate, $\gamma$ for the polarization decay rate of the atom, and $\kappa=\kappa_1+\kappa_{\rm loss}$ for the cavity field decay rate~\footnote{Note that our notation follows Ref.~\cite{Reiserer2015} with \(\gamma\) and \(\kappa\) denoting the amplitude decay rates. Another convention, used, e.g., in Ref.~\cite{Muniz2021}, uses the same symbols for the population decay rates (\(2\gamma\) and \(2\kappa\) in our notation). Therefore, using numbers from Ref.~\cite{Muniz2021}, we have to account for the factor of 2 difference.}, where $\kappa_1$ is the decay rate through the input mirror and $\kappa_{\rm loss}$ quantifies other cavity decay channels.
We consider a single-sided lossless cavity $\kappa_{\rm loss}=0$ in the regime $|\Delta|\sim\gamma\ll\kappa$, where the cavity response is approximately flat, while the atomic response varies sharply~\cite{Tiecke2014}.
In this limit, the standard input--output theory of a single-sided atom--cavity system gives the reflection amplitude
\begin{equation}\label{eq:reflection-amplitude}
    r(\Delta)=-\frac{1-2C+\ii\Delta/\gamma}{1+2C+\ii\Delta/\gamma},
\end{equation}
as derived in Methods.
Here $C=g_0^2/(2\kappa\gamma)$ is the atom--cavity cooperativity~\cite{Reiserer2015}.
The corresponding reflection probability and phase are
\begin{align}
    &R(\Delta)=\frac{(1-2C)^2+(\Delta/\gamma)^2}{(1+2C)^2+(\Delta/\gamma)^2},
    \label{eq:reflection-probability}\\
    &\phi_r(\Delta)=\pi+\arg(1-2C+\ii\Delta/\gamma)-\arg(1+2C+\ii\Delta/\gamma),
    \label{eq:reflection-phase}
\end{align}
with the phase restricted to the range $[-\pi,\pi]$.
The reflection probability exhibits a narrow dip centered on \(\Delta=0\), as shown in Fig.~\ref{fig:gedanken}(a).
The dip's full width at half maximum is $2(1+2C)\gamma\equiv\Gamma_{\rm P}$, equal to the Purcell-enhanced population decay rate, and the depth is $1-R(0)=8C/(1+2C)^2$.
The depth is maximal at the moderate cooperativity $C=0.5$, where for \(\Delta=0\) the prompt reflection from the input mirror and the field emitted back through the cavity port interfere destructively---the incident photon is removed from the reflected mode by atomic scattering~\cite{Borges2016}.
The same near-resonant structure produces a sharp phase response [Fig.~\ref{fig:gedanken}(b)].
Far from resonance, $r(\Delta)\to -1$, so $\phi_r$ approaches $\pm\pi$ for both positive and negative detunings.
Near resonance, the response depends qualitatively on the cooperativity: for $C<0.5$ the phase exhibits a $2\pi$ discontinuity, whereas at $C=0.5$ the phase changes by $\pi$ across a point where the reflected amplitude itself vanishes.
For $C>0.5$, the phase is continuous but steepens as $C$ approaches $0.5$ from above.
Near \(\Delta=0\), both the reflection probability and phase vary rapidly on the frequency scale set by \(\Gamma_P\).
Thus, a gravitational redshift comparable to or larger than this scale can produce substantial arm-to-arm differences in both quantities.

Notably, ultranarrow atomic resonances bring this regime within laboratory reach.
As a concrete example, consider the ${}^{87}\mathrm{Sr}$ clock transition, with $\omega_0\simeq2\pi\times429\,\mathrm{THz}$ and natural linewidth $2\gamma_0\simeq2\pi\times1.35\,\mathrm{mHz}$~\cite{Muniz2021}, coupled to a cavity with cooperativity \(C=0.5\).
In the lifetime-limited input--output convention, $\gamma=\gamma_0\simeq2\pi\times0.675\,\mathrm{mHz}$~\footnote{In an implementation with additional homogeneous broadening, $\gamma$ would be the measured polarization decay rate.}, and \(\Gamma_{\rm P}\simeq2\pi\times2.70\,{\rm mHz}\).
A height difference of $h=10\,\mathrm{cm}$ gives $|\delta\omega_g|\simeq2\pi\times4.68\,\mathrm{mHz}$, corresponding to $|\delta\omega_g|/\Gamma_{\rm P}\simeq1.73$.
The gravitational shift therefore exceeds the Purcell-enhanced response scale, and we can expect large differences between the arms already for centimetre-scale height differences.

To detect the effect of the gravitational field on a single-photon superposition state, we consider a height difference $h$ and a single photon impinging on the beam splitter with a detuning $\Delta_{\rm l}$ from the lower-arm scatterer.
We scan the interferometric phase $\varphi$ and record the photon-detection probability, which oscillates as $P_0(\varphi)=\tfrac{R_{\rm avg}}{2}\left[1-V\cos(\varphi+\Phi_g)\right]$, with $R_{\rm avg}=(R_{\rm l}+R_{\rm u})/2$.
The two observables are read directly from this fringe: the visibility $V$ from its amplitude, and the gravity-induced phase shift $\Phi_g$ from its position along $\varphi$ [Fig.~\ref{fig:gedanken}(c,d)].
The gravitational signal is revealed by scanning the probe detuning $\Delta_{\rm l}$ across the sharp resonance.
The experiment is then repeated for different height differences $h$.
In particular, at $h=0$, when the two scatterers experience the same gravitational potential, the two arms remain identical at every detuning, and neither the visibility nor the phase depends on $\Delta_{\rm l}$ [Fig.~\ref{fig:gedanken}(e,f), grey].
For $h\neq0$, instead, both quantities become modulated [Fig.~\ref{fig:gedanken}(e,f), black] as a result of the non-zero gravitational shift between the two scatterers.
Two settings are particularly instructive.
At $\Delta_{\rm l}=\omega_g\equiv|\delta\omega_g|$, the lower arm stays highly reflective while the redshift tunes the upper scatterer into resonance, suppressing its reflectivity.
The resulting amplitude imbalance encodes which-path information and reduces the fringe visibility.
The effect is largest at $C=0.5$, where the upper-arm reflectivity vanishes and $V\to0$ [Fig.~\ref{fig:gedanken}(d,e), orange], at the price of a reduced average reflection probability, $R_{\rm avg}\simeq0.46$.
In the complementary setting $\Delta_{\rm l}=-\Delta_{\rm u}=\omega_g/2$, the two arms are probed at equal and opposite detunings, so $R_{\rm l}=R_{\rm u}$ and the visibility remains unity, while the arms sample opposite sides of the dispersive response and acquire a large phase difference, $|\Phi_g|\simeq0.33\pi$ with $R_{\rm avg}\simeq0.75$ [Fig.~\ref{fig:gedanken}(d,f), green].
Importantly, the spectroscopic signal distinguishes the gravitational effect from other technical imperfections; for example, differential optical loss in the two arms, which would result in detuning-independent contrast reduction. 

\begin{figure}[t]
\includegraphics[width=\linewidth]{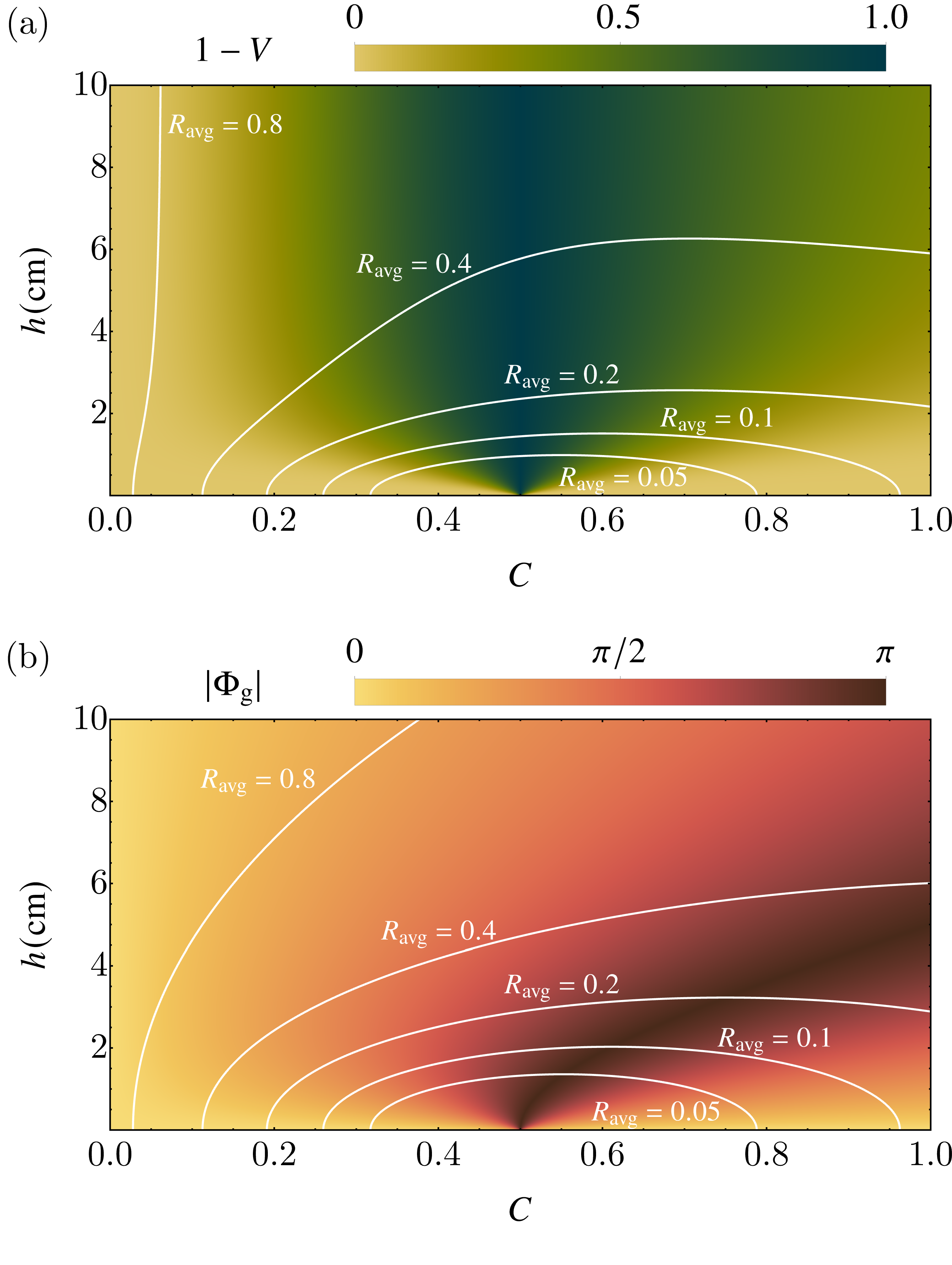}
\caption{\justifying
\textbf{Centimetre-scale gravitational redshifts produce large interferometric signals.}
(a) Visibility loss \(1-V\) for the \({}^{87}{\rm Sr}\) clock transition as a function of height separation \(h\) and atom--cavity cooperativity \(C\), evaluated at the upper-arm-resonance operating point, \(\Delta_{\rm u}=0\) and \(\Delta_{\rm l}=-\delta\omega_g\).
At \(C=0.5\), the reflected amplitude from the resonant arm vanishes, so the reflected photon carries complete which-path information, and the visibility is completely lost.
White contours show the average reflection probability \(R_{\rm avg}=(R_{\rm l}+R_{\rm u})/2\), which quantifies the fraction of photons contributing to the reflected-output signal.
(b) Gravity-induced phase difference \(|\Phi_g|\) evaluated at the symmetric-detuning operating point, \(\Delta_{\rm u}=\delta\omega_g/2\) and \(\Delta_{\rm l}=-\delta\omega_g/2\), for which \(R_{\rm l}=R_{\rm u}\) and the visibility remains unity.
}
\label{fig:visibility and phase}
\end{figure}
Figure~\ref{fig:visibility and phase} shows the resulting visibility loss and gravity-induced phase difference as functions of height and cooperativity, together with contours of the average reflection probability $R_{\rm avg}$. 
Remarkably, at $C=0.5$, the visibility loss is complete even for arbitrarily small height differences, since the operating point keeps the upper arm on resonance ($\Delta_{\rm u}=0$), where its reflected amplitude vanishes independently of $h$.
Note that for small height differences, this comes at the price of a strongly suppressed output field.
In particular, in the limit of $|\delta\omega_g|/\Gamma_{\rm P}\to0$, the output field vanishes, as $R_{\rm avg}\to0$.
To maintain significant output field amplitude, we need height differences corresponding to $|\delta\omega_g|\gtrsim\Gamma_{\rm P}$.
Importantly, the large effects do not rely on the exact reflection zero at $C=0.5$---visibility losses approaching unity and phase shifts of order $\pi$ occur over a broad cooperativity range.

\section*{Noise analysis and experimental requirements}
So far, our analysis assumes a monochromatic photon source, whereas a realistic probe necessarily has a finite bandwidth.
As shown in Methods, finite bandwidth can produce nonzero visibility loss even at symmetric detuning, where the monochromatic probe retains full visibility.
In fact, for \(h=10\,\mathrm{cm}\) and a Gaussian probe with linewidth \(\Delta\nu_{\rm FWHM}=10\,\mathrm{mHz}\)~\cite{Matei2017, Oelker2019}, symmetric detuning maximizes both the visibility loss and the phase magnitude, although at different cooperativities.
The visibility loss reaches \(1-V\simeq0.77\) at \(C\simeq1.10\), while the phase magnitude reaches \(|\Phi_g|\simeq0.50\pi\) at \(C\simeq1.86\).
Thus, substantial interferometric signatures persist at finite bandwidth.

A second practical issue is the stability and calibration of the two cavity-QED scatterers.
Since the signal is encoded in the relative scattering response of the two interferometer arms, common-mode fluctuations, such as probe-photon frequency noise or shifts shared by both scatterers, mainly displace the common operating point and do not, by themselves, generate path distinguishability or a differential phase between otherwise identical arms.
By contrast, differential fluctuations directly affect the relative scattering response of the two arms and can therefore contribute to the measured visibility and phase.
These include mismatches between the two scatterers in their atomic \(\delta\omega_{\rm a}\) or cavity \(\delta\omega_{\rm c}\) resonances, input-coupling \(\delta\kappa_1\) or excess-loss \(\delta\kappa_{\rm loss}\) rates, or atom--cavity coupling \(\delta g_0\), as well as scatterer-independent fractional amplitude-loss imbalance \(\delta\xi\) (as defined in Methods).
For each of these parameters, \(\delta p=p_{\rm u}-p_{\rm l}\) denotes the difference between the upper- and lower-arm scatterer.
Table~\ref{tab:noise-requirements} quantifies these channels: for each parameter, it gives the mismatch that alone produces a fraction \(\epsilon\) of the gravitational visibility loss.
Thus \(\epsilon=1\) corresponds to a mismatch large enough to reproduce the entire gravitational signal, and \(\epsilon=0.1\) to one contributing a ten-percent systematic error.
The differential atomic shift \(\delta\omega_{\rm a}\) is the most demanding channel in Table~\ref{tab:noise-requirements}.
It enters exactly as the gravitational detuning and is therefore indistinguishable from it in a static measurement.
For this reason, the measurement requires changing the height \(h\) and extracting the part of the signal that scales with it.
An independent spectroscopic calibration of the atomic resonances can further help to distinguish the signal from the noise.

\begin{table}[t]
\caption{\justifying
\textbf{Noise-control requirements for the \(^{87}{\rm Sr}\) implementation.}
The table gives the one-arm mismatch that produces a fraction \(\epsilon\) of the gravitationally induced visibility loss, \(1-V\simeq0.77\), corresponding to the \(h=10\,{\rm cm}\) height difference and \(10\,{\rm mHz}\) probe linewidth.
The reference cavity-QED system has \(\kappa=\kappa_1=2\pi\times70.4\,{\rm kHz}\), \(\kappa_{\rm loss}=0\), \(\gamma=2\pi\times0.675\,{\rm mHz}\), and \(g_0=2\pi\times10.2\,{\rm Hz}\), corresponding to \(C=1.1\).
The \(\epsilon=0.1\) column gives the scale for ten-percent systematic control, while \(\epsilon=1\) gives the mismatch required to mimic the full gravitational visibility-loss signal.
For the cavity-resonance mismatch, we report no \(\epsilon=1\) value, as it cannot reproduce the full gravitational signal on its own.
}
\label{tab:noise-requirements}
\begin{ruledtabular}
\begin{tabular}{lll}
Parameter & \(\epsilon=0.1\) & \(\epsilon=1\) \\
\hline
Atomic resonance, \(|\delta\omega_{\rm a}|/2\pi\) 
& \(\lesssim1.15\,{\rm mHz}\) 
& \(\lesssim4.68\,{\rm mHz}\) \\
Cavity resonance, \(|\delta\omega_{\rm c}|/2\pi\) 
& \(\lesssim53.0\,{\rm kHz}\) 
& non-limiting \\
Input coupling rate, \(|\delta\kappa_1|/2\pi\) 
& \(\lesssim27.6\,{\rm kHz}\) 
& \(\lesssim57.3\,{\rm kHz}\) \\
Excess loss rate, \(\delta\kappa_{\rm loss}/2\pi\) 
& \(\lesssim14.1\,{\rm kHz}\) 
& \(\lesssim55.6\,{\rm kHz}\) \\
Atom--cavity coupling, \(|\delta g_0|/2\pi\) 
& \(\lesssim2.69\,{\rm Hz}\) 
& \(\lesssim13.41\,{\rm Hz}\) \\
Scatterer-independent loss, \(\delta\xi\) 
& \(\lesssim0.333\) 
& \(\lesssim0.882\)
\end{tabular}
\end{ruledtabular}
\end{table}

Photon statistics impose a further constraint.
To remain in the weak-excitation regime, the incident photon flux must be smaller than the Purcell-enhanced decay rate, $\Gamma_{\rm P}\simeq2.71\times10^{-2}\,{\rm s}^{-1}$ at $C=1.1$, corresponding to a $37\,{\rm s}$ relaxation time.
Assuming the incoming photon flux $\dot{N}_{\rm in}=\Gamma_{\rm P}/5$, the shot-noise analysis in Methods yields a minimum integration time of approximately $4.5\,{\rm h}$ to resolve the reference visibility loss $1-V\simeq0.77$ with a signal-to-shot-noise ratio of three.
This estimate is a lower bound, as reference measurements, fringe tracking, background counts, and technical drifts introduce additional integration-time overhead.
The proposed scheme therefore trades the long propagation distances required by geometric proposals~\cite{Zych2011, Wu2024, Yu2026, Hilweg2017, Mohageg2022, Mohageg2025} for the long response time and low count rate imposed by the ultranarrow resonance.

Even with a weak coherent probe, the proposed experiment would provide a tabletop test of how Earth's gravitational redshift modifies optical interference---an effect that, to our knowledge, has not yet been directly observed.
A single-photon implementation would additionally probe the interplay between quantum mechanics and general relativity, provided that the probe field is independently certified as nonclassical.
Single photons at the \(^{87}\mathrm{Sr}\) clock-transition frequency, with a bandwidth narrow enough to resolve the scatterer response, could be generated using a third cavity-QED system~\cite{Hijlkema2007, Mucke2013} or by collecting the free-space emission of a tweezer-trapped atom with a high-numerical-aperture objective~\cite{Darquie2005}.
An extension to two-photon Hong--Ou--Mandel interference could instead probe gravitational effects on nonclassical photon correlations~\cite{Barzel2024, Gundogan2026}.
One could then certify nonclassicality within the interferometric measurement itself, without requiring separate source certification.

\section*{Conclusion and outlook}
The proposal highlights a complementary role of gravity in quantum interference.
In conventional matter-wave interferometric experiments, gravity is viewed primarily as a source of phase~\cite{Colella1975, Kasevich1992, Peters1999}.
Here, gravity also controls path distinguishability.
By shifting the photon frequency relative to an ultranarrow resonance, the gravitational field changes not only the phase acquired in each arm, but also the spectral state associated with each path.
The resulting loss of visibility is a manifestation of quantum complementarity: when the scattered optical field carries information about which arm the photon took, interference is reduced~\cite{Englert1996}.
In this sense, the cavity-QED scatterers act as spectral references that convert a relativistic frequency shift into measurable changes of phase and visibility.
Taken together, these results show that relativistic photonic interferometry need not rely exclusively on large propagation distances or long storage times.
Geometric accumulation can be replaced by spectral resolution near an ultranarrow optical resonance, allowing centimetre-scale height differences to produce large changes in optical scattering.
This opens a route to tabletop experiments in which gravitational time dilation modifies single-photon interference.

The mechanism proposed here is not tied to the particular strontium-based reflection scheme used in the analysis above.
The essential ingredient is an optical scattering amplitude that varies sharply on the scale of the gravitational frequency shift, or equivalently, an optical transition with a large quality factor \(Q=\omega_0/\gamma\).
Other optical-clock transitions, including other alkaline earth(-like) atoms or trapped ions~\cite{Ludlow2015}, could access the regime \(|\delta\omega_g|\sim\gamma\) on centimetre-to-tens-of-centimetres height scales.
The interferometer geometry is also flexible: a transmission-based Mach-Zehnder implementation, discussed in the Supplementary Material, realizes the same redshift-to-scattering principle with transmission amplitudes instead of reflection amplitudes.
More broadly, cavity-QED scatterers are just one implementation of a general idea: any sufficiently narrow and stable spectral feature, such as atomic resonances coupled to a waveguide~\cite{Vetsch2010, Goban2014, Sheremet2023},
electromagnetically~\cite{Boller1991, Hau1999, Fleischhauer2005} or optomechanically~\cite{Weis2010, Yuan2015} induced transparency resonance, could convert a gravitational redshift into path-dependent optical response.

\section*{Methods}
\subsection*{Interferometric visibility and phase}
We derive the interferometric observables using a single-photon path-state description.
We denote the photon states in the lower and upper arms by $\ket{l}$ and $\ket{u}$, respectively.
After the first balanced beam splitter, the photon is prepared in the path superposition $\ket{\psi_1}=(\ket{l}+\ii\ket{u})/\sqrt{2}$, where the phase factor $\ii$ follows from our beam-splitter convention.
The photon is then reflected from the cavity-QED scatterers in the two arms, with complex amplitudes
\begin{equation}
    r_{\rm l,u}=\sqrt{R_{\rm l,u}}\exp(\ii\phi_{\rm l,u}),
\end{equation}
while an externally controlled phase $\varphi$ is applied to the upper arm.
After recombination at the beam splitter, the unnormalised state in the two interferometer output modes is
\begin{equation}\label{final state; monochromatic}
    \ket{\psi_f}=\frac{1}{2}\left[(r_{\rm l}-\ee^{\ii\varphi}r_{\rm u})\ket{0}+\ii(r_{\rm l}+\ee^{\ii\varphi}r_{\rm u})\ket{1}\right],
\end{equation}
where $\ket{0}$ and $\ket{1}$ denote the two output ports.
The state $\ket{\psi_f}$ is unnormalised, $\braket{\psi_f}{\psi_f}\leq 1$, because the cavity-QED scatterers can remove photons from the reflected interferometer modes.
Its norm,
\begin{equation}
    \braket{\psi_f}{\psi_f}=\frac{R_{\rm l}+R_{\rm u}}{2}\equiv R_{\rm avg},
\end{equation}
quantifies the total reflected-output signal.
We define the complex interference amplitude
\begin{equation}
    \mathcal{A}=r_{\rm l}^\ast r_{\rm u}.
\end{equation}
Then, the probability of detecting the photon in the output port $i\in\{0,1\}$ is given by
\begin{equation}
    P_i(\varphi)=\frac{R_{\rm avg}}{2}\left[1-(-1)^i\frac{|\mathcal{A}|}{R_{\rm avg}}\cos\left(\varphi+\arg\mathcal{A}\right)\right].
\end{equation}
We identify
\begin{equation}\label{methods: visibility and phase}
    V=\frac{|\mathcal{A}|}{R_{\rm avg}},\qquad \Phi_g=\arg\mathcal{A},
\end{equation}
as the fringe visibility and the gravity-induced phase difference, respectively.
By noting that $|\mathcal{A}|=\sqrt{R_{\rm l} R_{\rm u}}$ and $\arg\mathcal{A}=\phi_{\rm u}-\phi_{\rm l}$, we recover the formulas from the main text [see Eq.~\eqref{visibility and phase}].
\subsection*{Cavity-QED reflection amplitude}
We derive here the optical reflection amplitude used to describe the cavity-QED scatterers in the main text.
We use the standard weak-excitation reflection amplitude obtained from input--output theory~\cite{Gardiner1985} for a single cavity mode coupled to a single atomic transition~\cite{Hu2008, Tiecke2014}.
For a monochromatic component of the incident field with frequency \(\omega\), the reflection amplitude reads~\cite{Hu2008, Tiecke2014, Reiserer2015}
\begin{equation}
    r(\omega)
    =
    1-
    \frac{2\kappa_1\left(\ii\Delta_{\rm a}+\gamma\right)}
    {\left(\ii\Delta_{\rm c}+\kappa\right)\left(\ii\Delta_{\rm a}+\gamma\right)+g_0^2}.
    \label{general-reflection-amplitude}
\end{equation}
Here \(\Delta_{\rm a}=\omega-\omega_{\rm a}\) and \(\Delta_{\rm c}=\omega-\omega_{\rm c}\) are the detunings from the atomic transition frequency \(\omega_{\rm a}\) and the cavity resonance frequency \(\omega_{\rm c}\), respectively; \(\gamma\) denotes the polarization decay rate of the atom, \(\kappa_1\) is the coupling rate through the input mirror, and \(\kappa_{\rm loss}=\kappa-\kappa_1\) collects all other cavity decay channels, such as transmission through other ports, scattering, or absorption.
It is useful to rewrite Eq.~\eqref{general-reflection-amplitude} in terms of dimensionless detunings $\chi_{\rm a}=\Delta_{\rm a}/\gamma$, $\chi_{\rm c}=\Delta_{\rm c}/\kappa$, the atom--cavity cooperativity $C=g_0^2/(2\kappa\gamma)$, and the input-port decay fraction $\eta=\kappa_1/\kappa$.
The reflection amplitude then becomes
\begin{equation}\label{reflection-rescaled}
    r(\chi_{\rm a},\chi_{\rm c})=1-2\eta\frac{1+\ii\chi_{\rm a}}{(1+\ii\chi_{\rm c})(1+\ii\chi_{\rm a})+2C}.
\end{equation}
We focus on the regime in which the atomic polarization decays much more slowly than the cavity field, \(\gamma\ll\kappa\).
The gravitational frequency shift can be comparable to the first, but remains much smaller than the latter, $|\delta\omega_g|\sim\gamma\ll\kappa$.
Consequently, we can neglect the gravity-induced difference in cavity detuning, approximating \(\chi_{\rm c}\simeq{\rm const}\) over the relevant frequency range.
We further assume that both the atomic transition and the probe photons are tuned to the cavity resonance, so that $\Delta_{\rm a},\omega_{\rm a}-\omega_{\rm c}\ll\kappa$, and consequently, set $\chi_{\rm c}\simeq0$.
Therefore, the remaining frequency dependence is described by the photon-atom detuning $\chi_{\rm a}$.
Finally, for an ideal single-sided lossless cavity, $\eta=1$, Eq.~\eqref{reflection-rescaled} reduces to
\begin{equation}
r(\chi_{\rm a})=-\frac{1-2C+\ii\chi_{\rm a}}{1+2C+\ii\chi_{\rm a}},
\end{equation}
which is the expression used in the main text [see Eq.~\eqref{eq:reflection-amplitude}].
\subsection*{Small redshift regime}
We consider the perturbative regime in which the gravitationally induced detuning difference, $\delta\chi_{\rm a}^{(g)}=\delta\omega_g/\gamma$, is small enough that the two arms probe nearly the same reflection amplitude.
Writing $r_{\rm u}=r_{\rm l}+\delta r$, this requires $|\delta r|\ll|r_{\rm l}|$.
Since $\delta r=r'(\chi_{\rm a,l})\delta\chi_{\rm a}^{(g)}$, the expansion is valid provided $\left|\delta\chi_{\rm a}^{(g)}\right|\ll|r(\chi_{\rm a,l})/r'(\chi_{\rm a,l})|$.
Expanding the visibility and phase difference to the lowest non-vanishing order in $\delta\chi_{\rm a}^{(g)}$, we obtain
\begin{equation}
\begin{split}\label{small redshift expansion}
    &1-V\simeq\frac{32C^2\chi_{\rm a,l}^2\left(\delta\chi_{\rm a}^{(g)}\right)^2}{(((1-2C)^2+\chi_{\rm a,l}^2)((1+2C)^2+\chi_{\rm a,l}^2))^2},\\
    &\Phi_g\simeq\frac{4C(1-4C^2-\chi_{\rm a,l}^2)\delta\chi_{\rm a}^{(g)}}{((1-2C)^2+\chi_{\rm a,l}^2)((1+2C)^2+\chi_{\rm a,l}^2)}.
\end{split}
\end{equation}
Thus, in the perturbative regime, the visibility loss is quadratic in the gravity-induced detuning, whereas the phase response is linear.
%Small redshifts are therefore more naturally accessed through phase readout than through contrast loss: resolving the latter requires sensitivity to a second-order change in the fringe visibility, while phase difference appears already in the first order.
The dimensionless redshift parameter can be written as
\begin{equation}
    \delta\chi_{\rm a}^{(g)}=\frac{\delta\omega_g}{\gamma}\simeq-\frac{\omega_{\rm l}}{\gamma}\frac{gh}{c^2}\simeq-\frac{\omega_{\rm a}}{\gamma}\frac{gh}{c^2}=-Q\frac{gh}{c^2},
\end{equation}
where in the last two steps we used $\omega_{\rm l}\simeq\omega_{\rm a}$ and introduced the atomic quality factor $Q=\omega_{\rm a}/\gamma$.
Equation~\eqref{small redshift expansion} therefore reproduces the scaling $\Phi_g\sim Qgh/c^2$ quoted in the main text [see Eq.~\eqref{phase with scatterers}], with a dimensionless prefactor set by cooperativity and operating detuning.
The magnitude of this prefactor is largest at resonance, $\chi_{\rm a,l}=0$, where the phase response reduces to
\begin{equation}
    \Phi_g\simeq\frac{4C}{1-4C^2}\delta\chi_{\rm a}^{(g)}.
\end{equation}
For $C\sim1$, the prefactor is of order unity; it vanishes in the limits $C\to0$ and $C\to\infty$, and formally diverges as $C\to0.5$.
This divergence does not imply an arbitrarily large measurable phase shift: at $C=0.5$ and $\chi_{\rm a,l}=0$, the reflected amplitude vanishes, $r_{\rm l}=0$, so the reflected phase is undefined and the perturbative expansion breaks down.
\subsection*{Finite-bandwidth probe field}
The main-text analysis assumes a monochromatic input field.
This approximation captures the basic mechanism by which a gravitational redshift is converted into a difference between the scattering amplitudes in the two interferometer arms. However, for ultranarrow optical transitions, the probe-field spectral width becomes an important parameter.
If the probe bandwidth is comparable to or larger than the frequency range over which the cavity-QED reflection amplitude varies, different spectral components acquire distinct amplitudes and phases.
Here, we describe the input photon by a normalized spectral envelope $f(\omega)$, with
\begin{equation}
    \int\dd\omega\,|f(\omega)|^2=1.
\end{equation}
The state at the interferometer output is given by
\begin{equation}
    \begin{split}
        \ket{\psi_f}=\frac{1}{2}\int\dd\omega\,f(\omega)\big\{&[r_{\rm l}(\omega)-\ee^{\ii\varphi}r_{\rm u}(\omega)]\ket{0,\omega}\\
        &+\ii[r_{\rm l}(\omega)+\ee^{\ii\varphi}r_{\rm u}(\omega)]\ket{1,\omega}\big\},
    \end{split}
\end{equation}
i.e., the path degree of freedom is now accompanied by a spectral degree of freedom $\ket{\omega}$, and the reflection amplitudes acquire an additional $\omega$-dependence (compare with Eq.~\eqref{final state; monochromatic}).
Referring to the notation used in the previous subsections, we have $r_{\rm l}(\omega)\equiv r(\omega)$ and $r_{\rm u}(\omega)\equiv r(\omega+\delta\omega_g)$.
A frequency non-resolving measurement of the output probabilities is described by a positive operator-valued measure (POVM) $\{\Pi_i\}_{i\in\{0,1\}}$ with
\begin{equation}
    \Pi_i=\int\dd\omega\,\ket{i,\omega}\bra{i,\omega},
\end{equation}
leading to the probability of the $i$-th outcome, $P_i=\bra{\psi_f}\Pi_i\ket{\psi_f}$, given by
\begin{equation}
    P_i(\varphi)=\frac{1}{4}\int\dd\omega\,|f(\omega)|^2\left|r_{\rm l}(\omega)-(-1)^i\ee^{\ii\varphi}r_{\rm u}(\omega)\right|^2.
\end{equation}
We introduce the effective (frequency-averaged) quantities
\begin{equation}\label{effective quantities}
    \begin{split}
        &R_{\rm l,u}^{\rm eff}=\int\dd\omega\,|f(\omega)|^2|r_{\rm l,u}(\omega)|^2,\\
        &R_{\rm avg}^{\rm eff}=\frac{R_{\rm l}^{\rm eff}+R_{\rm u}^{\rm eff}}{2},\\
        &\mathcal{A}^{\rm eff}=\int\dd\omega\,|f(\omega)|^2r_{\rm l}^\ast(\omega)r_{\rm u}(\omega).
    \end{split}
\end{equation}
This allows us to rewrite the probability of the $i$-th outcome in the following compact form:
\begin{equation}
    P_i(\varphi)=\frac{R_{\rm avg}^{\rm eff}}{2}\left[1-(-1)^i\frac{|\mathcal{A}^{\rm eff}|}{R_{\rm avg}^{\rm eff}}\cos\left(\varphi+\arg\mathcal{A}^{\rm eff}\right)\right].
\end{equation}
Here, we identify the effective visibility and phase
\begin{equation}\label{finite-bandwidth observables}
    V^{\rm eff}=\frac{|\mathcal{A}^{\rm eff}|}{R_{\rm avg}^{\rm eff}},\qquad \Phi_g^{\rm eff}=\arg\mathcal{A}^{\rm eff},
\end{equation}
which generalizes Eq.~\eqref{methods: visibility and phase} derived for a monochromatic input.

\begin{figure}[h!]
\includegraphics[width=\linewidth]{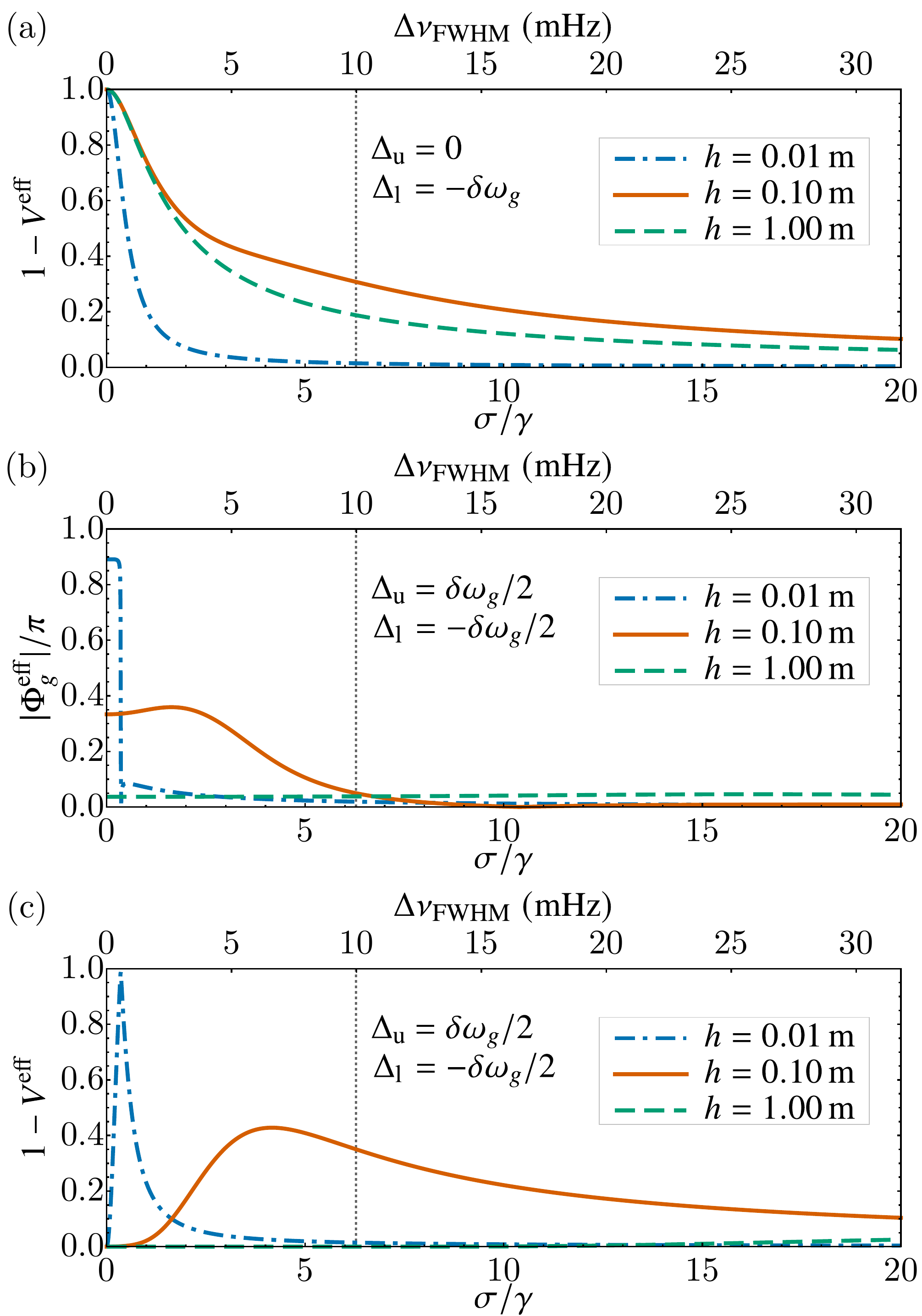}
\caption{\justifying \textbf{Finite-bandwidth response of the interferometric signatures.}
Visibility and phase response for a Gaussian probe spectrum with angular-frequency bandwidth \(\sigma\).
(a) Effective visibility loss \(1-V^{\rm eff}\) for the upper-arm-resonance operating point, \(\Delta_{\rm u}=0\) and \(\Delta_{\rm l}=-\delta\omega_g\).
(b) Effective gravity-induced phase shift \(|\Phi_g^{\rm eff}|/\pi\) for the symmetric-detuning operating point, \(\Delta_{\rm u}=\delta\omega_g/2\) and \(\Delta_{\rm l}=-\delta\omega_g/2\).
(c) Effective visibility loss for the symmetric detuning.
The curves correspond to different vertical separations \(h\) for the \({}^{87}{\rm Sr}\) clock transition and \(C=0.5\).
The lower horizontal axis gives the dimensionless bandwidth \(\sigma/\gamma\), while the upper axis gives the corresponding Gaussian full width at half maximum, \(\Delta\nu_{\rm FWHM}=\sqrt{2\ln2}\,\sigma/\pi\).
The vertical dashed line marks \(\Delta\nu_{\rm FWHM}=10\,{\rm mHz}\), corresponding to
\(\sigma/\gamma\simeq6.29\) for \(\gamma/2\pi=0.675\,{\rm mHz}\).
\label{fig:finite-bandwidth}}
\end{figure}

For the numerical estimates shown in Fig.~\ref{fig:finite-bandwidth}, we model the finite spectral width of the probe by a Gaussian frequency distribution,
\begin{equation}\label{eq:probe-frequency-distribution}
|f(\omega)|^2=\frac{1}{\sqrt{2\pi}\sigma}\exp\left[-\frac{(\omega-\bar{\omega})^2}{2\sigma^2}\right],
\end{equation}
where \(\bar{\omega}=\omega_0+\Delta_{\rm l}\) is the central probe frequency (shifted by \(\Delta_{\rm l}\) from the atom--cavity resonance \(\omega_0\)) and \(\sigma\) is the bandwidth in angular-frequency units.
For this Gaussian spectrum, the frequency averages in Eq.~\eqref{effective quantities} can be evaluated analytically, yielding closed-form expressions in terms of the Faddeeva function~\cite{DLMF} (see Supplementary Material).
These expressions recover the monochromatic limit for \(\sigma/\gamma\ll1+2C\), corresponding to a probe spectrum much narrower than the (Purcell-broadened) atomic linewidth.
In contrast, \(\sigma/\gamma\gtrsim1+2C\) corresponds to a probe whose bandwidth resolves a non-negligible part of the cavity-QED scattering response.
The bandwidth \(\sigma\) is related to the full width at half maximum, $\Delta\nu_{\rm FWHM}$, of the corresponding Gaussian spectrum by
\begin{equation}
\sigma=2\pi\,\frac{\Delta\nu_{\rm FWHM}}{2\sqrt{2\ln 2}}.
\end{equation}
In Fig.~\ref{fig:finite-bandwidth}, we mark a Gaussian probe linewidth \(\Delta\nu_{\rm FWHM}=10\,{\rm mHz}\), corresponding to \(\sigma/\gamma\simeq6.29\), for comparison with state-of-the-art ultrastable lasers based on cryogenic silicon reference cavities~\cite{Matei2017, Oelker2019}.

The gravity-induced visibility loss at the upper-arm-resonance point \(\Delta_{\rm u}=0\) decreases with increasing bandwidth \(\sigma\) [Fig.~\ref{fig:finite-bandwidth}(a)], but remains substantial at \(10\,{\rm mHz}\).
For \(h=10\,{\rm cm}\), it is \(1-V^{\rm eff}\simeq0.31\).
At fixed bandwidth, however, the loss need not increase monotonically with height.
For example, it is larger for \(h=10\,{\rm cm}\) than for \(h=1\,{\rm m}\), where \(1-V^{\rm eff}\simeq0.19\).
This behavior is determined by the relative detuning of the photons from the narrow reflection dip.
When the gravitational shift is comparable to the probe bandwidth, both wavepackets overlap the narrow reflection dip, which attenuates different parts of their spectra.
This produces strongly mismatched reflected wavepackets, making them distinguishable, thus reducing the visibility.
When \(|\delta\omega_g|\gg\sigma\), by contrast, the photon wavepacket in the lower arm is far off-resonance from the reflection dip and effectively probes a flat-response region of the mirror, so it is reflected without appreciable attenuation or spectral distortion.
The two wavepackets can then exhibit greater spectral overlap than in the intermediate-shift regime.
Consequently, the visibility loss is governed by the redshift relative to the probe bandwidth, rather than by the height difference alone.

The gravity-induced phase difference at symmetric detuning is more sensitive to the probe bandwidth [Fig.~\ref{fig:finite-bandwidth}(b)].
For \(h=10\,{\rm cm}\), it is \(|\Phi_g^{\rm eff}|\simeq0.33\pi\) in the monochromatic limit and initially increases to about \(0.36\pi\) for a linewidth near \(2.6\,{\rm mHz}\).
At larger bandwidths, spectral averaging reduces the phase response---at \(10\,{\rm mHz}\), it remains \(|\Phi_g^{\rm eff}|\simeq0.050\pi\), or about \(0.16\) rad.
Thus a finite linewidth does not rule out phase readout, but it requires a phase uncertainty well below \(0.05\pi\).
A probe linewidth of a few millihertz instead retains a phase shift of order \(0.3\pi\).
Notably, for a finite probe bandwidth, the visibility loss does not vanish at symmetric detuning [Fig.~\ref{fig:finite-bandwidth}(c)], unlike in the monochromatic limit.
Although the two central detunings correspond to equal reflection probabilities, the finite-bandwidth wavepackets sample opposite sides of the reflection dip.
Consequently, the two scatterers attenuate different parts of the wavepackets' spectra, so they are not identical at recombination, and their overlap remains below unity.
At \(h=10\,{\rm cm}\) and \(10\,{\rm mHz}\), the resulting loss is \(1-V^{\rm eff}\simeq0.35\), slightly larger than the upper-arm-resonance value.

\begin{figure}[ht!]
\includegraphics[width=\linewidth]{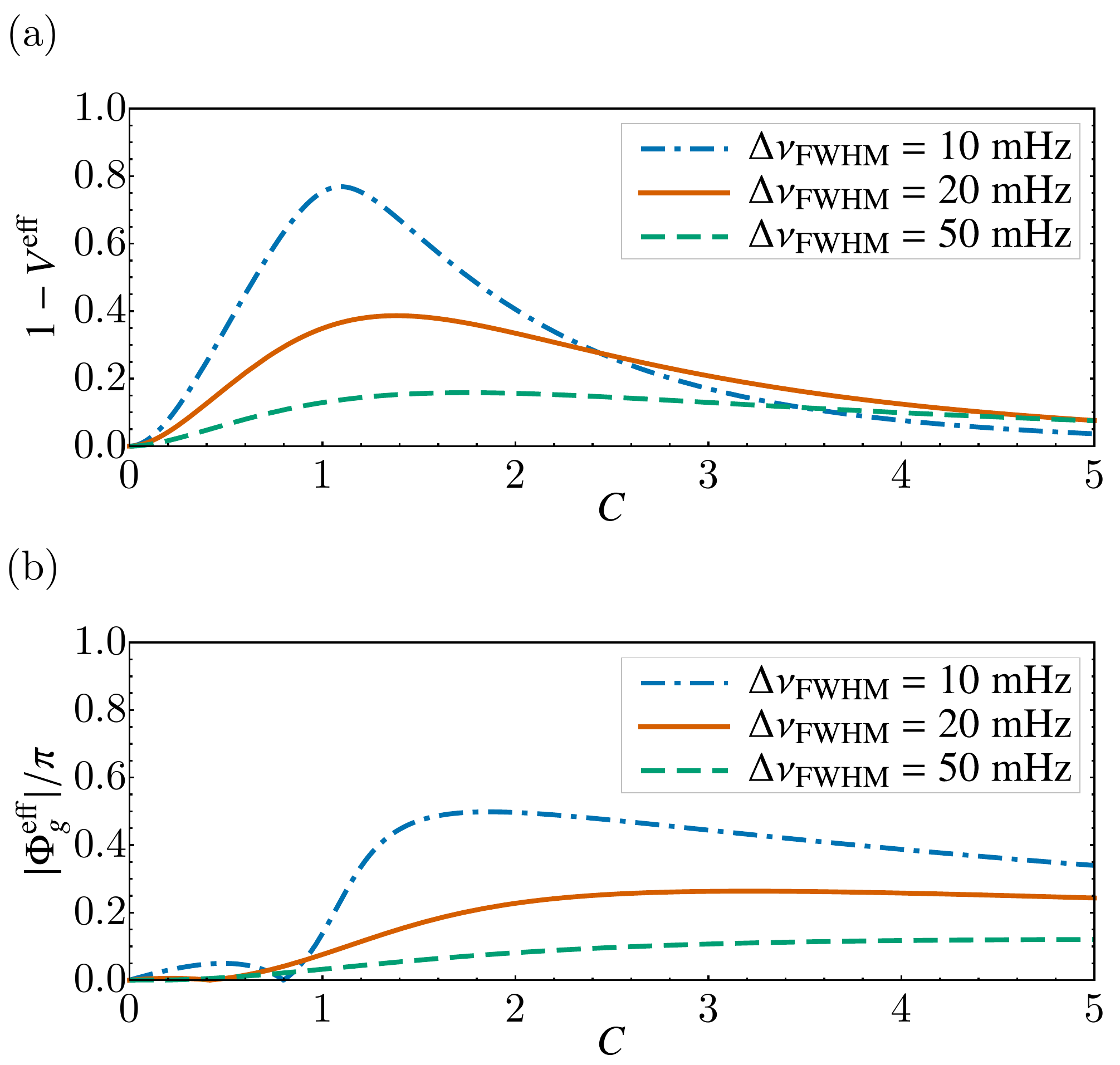}
\caption{\justifying \textbf{Interferometric response versus cooperativity at symmetric detuning.}
(a) Effective visibility loss \(1-V^{\rm eff}\) and (b) effective gravity-induced phase shift \(|\Phi_g^{\rm eff}|/\pi\) versus atom--cavity cooperativity \(C\), for probe linewidths \(\Delta\nu_{\rm FWHM}=10\), \(20\), and \(50\,\mathrm{mHz}\).
The curves are calculated for \(h=10\,\mathrm{cm}\) and symmetric detuning, \(\Delta_{\rm u}=-\Delta_{\rm l}=\delta\omega_g/2\).
\label{fig:finite-bandwidth-cooperativity-optimization}}
\end{figure}
Interestingly, at the finite probe bandwidth considered here, \(C=0.5\) does not maximize either interferometric signature.
For \(h=10\,\mathrm{cm}\) and a Gaussian probe with \(\Delta\nu_{\rm FWHM}=10\,\mathrm{mHz}\), the numerical search finds both maxima at symmetric detuning, but at different cooperativities.
The visibility loss reaches \(1-V^{\rm eff}\simeq0.77\) at \(C\simeq1.10\), with \(R_{\rm avg}^{\rm eff}\simeq0.66\); the phase magnitude reaches \(|\Phi_g^{\rm eff}|\simeq0.50\pi\) at \(C\simeq1.86\), with \(R_{\rm avg}^{\rm eff}\simeq0.65\).
Figure~\ref{fig:finite-bandwidth-cooperativity-optimization} shows both signatures as functions of \(C\) at symmetric detuning and for different probe linewidths.
We observe that increasing linewidth decreases the maximal achievable effects, while the optimal cooperativity, at which these maximal effects are obtained, increases.
We use the visibility-loss optimum at \(C\simeq1.10\) as the reference for the noise analysis below.

\subsection*{Technical imperfections and control requirements}
In this section, we formulate a minimal noise model for imperfections in the two cavity-QED scatterers.
Since the gravitational signal is encoded in the relative scattering response of the two interferometer arms, differential technical fluctuations can imitate or obscure the gravitationally induced visibility loss and phase shift.
A complete noise budget depends on the specific cavity-QED implementation.
Here we take a platform-independent approach: we assume that the complex scattering response of each arm has been independently characterized near the operating point, and we use this response to estimate the level of parameter mismatch that would produce a visibility loss comparable to the gravitational signal.
For each arm \(j\in\{l,u\}\), we write the reflection amplitude as
\begin{equation}
    r_j=r\left(\omega_{\rm a}^{(j)},\omega_{\rm c}^{(j)},\kappa_1^{(j)},\kappa_{\rm loss}^{(j)},g_0^{(j)}\right),
\end{equation}
with \(r\) given by Eq.~\eqref{general-reflection-amplitude}.
In addition to the parameters of each scatterer, we consider scatterer-independent losses in each arm by rescaling \(r_j\to\left(1-\xi^{(j)}\right)r_j\), where \(\xi^{(j)}\in[0,1]\) quantifies the loss in the \(j\)-th arm.
We treat
\begin{equation}
    \omega_{\rm a}^{(j)},\quad \omega_{\rm c}^{(j)},\quad \kappa_1^{(j)},\quad \kappa_{\rm loss}^{(j)},\quad g_0^{(j)},\quad \xi^{(j)}
\end{equation}
as the primitive parameters whose fluctuations can modify the interferometric response.
The atomic polarization decay rate \(\gamma\) is kept fixed as a calibrated property of the chosen transition.
% The numerical estimates use the lifetime-limited value \(\gamma=\Gamma_0/2\).
% Differential fluctuations of the linewidth are not included in this minimal noise model.
For any parameter \(p^{(j)}\in\{\omega_{\rm a}^{(j)},\omega_{\rm c}^{(j)},\kappa_1^{(j)}, \kappa_{\rm loss}^{(j)},g_0^{(j)},\xi^{(j)}\}\), we decompose fluctuations about a reference operating point \(p_0\) into common- and differential-mode components,
\begin{equation}
    \begin{split}
        p_{\rm com}(t)&=\frac{p_{\rm u}(t)+p_{\rm l}(t)}{2}-p_0,\\
        p_{\rm diff}(t)&=p_{\rm u}(t)-p_{\rm l}(t).
    \end{split}
\end{equation}
% Equivalently,
% \begin{equation}
%     \begin{split}
%         p_{\rm u}(t)&=p_0+p_{\rm com}(t)+\frac{1}{2}p_{\rm diff}(t),\\
%         p_{\rm l}(t)&=p_0+p_{\rm com}(t)-\frac{1}{2}p_{\rm diff}(t).
%     \end{split}
% \end{equation}
The effect of these fluctuations on the measurement can be expressed in terms of the two interferometric observables considered in the main text: the visibility \(V\) and the differential phase \(\Phi_g\).
In the monochromatic case (see Eq.~\eqref{methods: visibility and phase}),
\begin{equation}
    V=\frac{2|r_{\rm l}||r_{\rm u}|}{|r_{\rm l}|^2+|r_{\rm u}|^2},
    \qquad
    \Phi_g=\arg(r_{\rm l}^\ast r_{\rm u}).
\end{equation}
At the symmetric reference point, where the gravitational redshift and differential fluctuations are absent, the two scatterers are identical, \(r_{\rm u}=r_{\rm l}\).
Provided the reflected signal is nonzero, this gives
\begin{equation}
    \left.V\right|_{r_{\rm u}=r_{\rm l}}=1,
    \qquad
    \left.\Phi_g\right|_{r_{\rm u}=r_{\rm l}}=0,
\end{equation}
independently of the value of the common-mode displacement.
Thus, common-mode fluctuations do not, by themselves, create path distinguishability or a differential phase between otherwise identical arms.
The same conclusion holds for a finite-bandwidth probe after replacing \(\mathcal{A}\), \(R_{\rm avg}\), \(V\), and \(\Phi_g\) by their effective, frequency-averaged counterparts.
At the symmetric reference point, a common-mode fluctuation changes the two scattering amplitudes in the same way, \(r_{\rm l}(\omega)=r_{\rm u}(\omega)\), for every spectral component.
Hence \(V^{\rm eff}=1\) and \(\Phi_g^{\rm eff}=0\), again provided the total reflected signal is nonzero.
Common-mode fluctuations can still change the total reflected signal, \(R_{\rm avg}\) or \(R_{\rm avg}^{\rm eff}\), and can shift the local susceptibility of the interferometer to a subsequent gravitational or differential perturbation.
By contrast, differential-mode fluctuations change the relative scattering response of the two arms and can therefore imitate, obscure, or add to the effect of a gravitational redshift.

We now quantify how large a technical mismatch must be to produce a visibility loss comparable to that generated by the gravitational redshift.
We focus on the finite-bandwidth visibility loss \(L = 1 - V^{\rm eff}\), with \(V^{\rm eff}\) defined in Eq.~\eqref{finite-bandwidth observables}.
Although the common- and differential-mode decomposition above is useful for classifying fluctuations, the numerical requirements are formulated as one-arm mismatches.
Specifically, the lower-arm scatterer is kept at its calibrated reference value, while one parameter of the upper-arm scatterer is varied,
\begin{equation}
    p_{\rm l}=p_0,\qquad p_{\rm u}=p_0+\delta p .
\end{equation}
Thus \(\delta p=p_{\rm u}-p_{\rm l}\) is the differential mismatch between the two arms.
Let
\begin{equation}
    L_g = 1 - V^{\rm eff}_g
\end{equation}
denote the visibility loss produced by the target gravitational redshift in the absence of technical noise.
For a given primitive parameter \(p\), we define
\begin{equation}
    L_p(\delta p) = 1 - V^{\rm eff}_p(\delta p),
\end{equation}
where \(V^{\rm eff}_p(\delta p)\) is the finite-bandwidth visibility when the gravitational redshift is set to zero, and the only imperfection is a one-arm mismatch \(\delta p\) of parameter \(p\).
The control requirement at fractional level \(\epsilon\) is then defined by
\begin{equation}\label{eq:epsilon-control-requirement}
    L_p\left(\delta p_{\rm req}^{(\epsilon)}\right) = \epsilon L_g .
\end{equation}
The case \(\epsilon=1\) gives the mismatch required to reproduce the full gravitationally induced visibility loss, while \(\epsilon=0.1\) gives the scale at which the mismatch contributes ten percent of the gravitational signal.

The numerical requirements reported in Table~\ref{tab:noise-requirements} are evaluated for the \(^{87}{\rm Sr}\) clock transition with \(\gamma/2\pi=0.675\,{\rm mHz}\) and a Gaussian probe spectrum with \(\Delta\nu_{\rm FWHM}=10\,{\rm mHz}\), equivalently \(\sigma/\gamma=6.29\).
The reference parameters are set as follows.
The cavity is assumed to be lossless, \(\kappa_{\rm loss}=0\), and the coupling through the input mirror is chosen to match the cavity linewidth from Ref.~\cite{Muniz2021}, \(\kappa_1=2\pi\times70.4\,{\rm kHz}\).
Further, we assume that, in the reference setting, the atom is resonant with the cavity, \(\omega_{\rm a}=\omega_{\rm c}\), the coupling strength is \(g_0=2\pi\times10.2\,{\rm Hz}\), resulting in \(C=1.1\), and there are no additional scatterer-independent losses, \(\xi=0\). For the reference value of the gravity-induced visibility loss, we take \(L_g=0.767\), corresponding to a height difference \(h=10\,{\rm cm}\) and symmetric detuning \(\Delta_{\rm u}=-\Delta_{\rm l}=\delta\omega_g/2\).
%Scanning the central frequency of the probe, as defined in Eq.~\eqref{eq:probe-frequency-distribution}, gives the maximum visibility loss at symmetric detuning.
For each parameter \(p\), we scan over the physically allowed one-arm mismatches \(\delta p\), such that the upper-arm value \(p_{\rm u}=p_0+\delta p\) remains in the physical domain, and over the central frequency of the probe.
We determine \(\delta p_{\rm req}^{(\epsilon)}\) as the smallest (in terms of magnitude) mismatch satisfying Eq.~\eqref{eq:epsilon-control-requirement} in the scanned range.
If Eq.~\eqref{eq:epsilon-control-requirement} has no solution, the corresponding channel is reported in Table~\ref{tab:noise-requirements} as non-limiting at that value of \(\epsilon\).
In particular, this is the case for \(\epsilon=1\) and the cavity-resonance mismatch \(\delta\omega_{\rm c}\)---the maximal visibility loss obtained by varying the latter parameter is given by
\begin{equation}
    \max_{\delta\omega_{\rm c}}L_{\omega_{\rm c}}(\delta\omega_{\rm c})\simeq 0.518 < L_g,
\end{equation}
which is reached for $|\delta\omega_{\rm c}|\gg\kappa$.
% For stochastic fluctuations, the relevant rms mismatch is obtained after filtering by the measurement protocol.
% The requirements in Table~\ref{tab:noise-requirements} should then be understood as upper bounds on this rms mismatch.

\subsection*{Shot noise and integration time}
The reflection amplitude in Eq.~\eqref{general-reflection-amplitude}, used throughout this work, is valid only in the weak-excitation regime.
This condition limits the usable input photon flux, thereby making photon-counting shot noise one of the central experimental constraints of the proposed setup.
For a sequence of single-photon probes, the weak-excitation condition requires the photon flux entering the interferometer to be smaller than the Purcell-enhanced population decay rate of the atom,
\begin{equation}\label{eq:input photon flux}
    \dot{N}_{\rm in}=\alpha\Gamma_{\rm P},
\end{equation}
where the parameter $\alpha\lesssim1$ provides a safety margin against saturation.
Reducing \(\alpha\) improves the validity of the weak-excitation approximation but increases the integration time required to overcome photon-counting shot noise.

We assume we measure the single output port for equal times at the maximum and minimum of the interference fringe, yielding counts $n_{\max}$ and $n_{\min}$.
For a simple shot noise estimate, we model these counts as independent Poisson variables with mean values
\begin{equation}
    \overline{n}_{\max,\min}=\frac{\dot N_{\rm in}T}{4}\,\beta R_{\rm avg}^{\rm eff}\left(1\pm V^{\rm eff}\right),
\end{equation}
where $T$ is the total acquisition time and $\beta$ contains the propagation and detection efficiencies not included in $R_{\rm avg}^{\rm eff}$.
Conditioning on the total number of counts $N_{\rm det}=n_{\max}+n_{\min}$, we can treat $n_{\max}$ as a binomially distributed random variable
\begin{equation}
    n_{\max}\sim{\rm B}
    \left(N_{\rm det},\frac{1+V^{\rm eff}}{2}\right)
\end{equation}
with variance
\begin{equation}
    \sigma_{n_{\max}}^2=\frac{N_{\rm det}}{4}\left(1-(V^{\rm eff})^2\right)
\end{equation}
associated with shot noise.
Consequently, the estimator
\begin{equation}
    \widehat V=\frac{n_{\max}-n_{\min}}{n_{\max}+n_{\min}}=\frac{2n_{\max}}{N_{\rm det}}-1
\end{equation}
has the shot-noise-induced variance
\begin{equation}
    \sigma_V^2=\frac{4\sigma_{n_{\max}}^2}{N_{\rm det}^2}=\frac{1-(V^{\rm eff})^2}{N_{\rm det}}.
\end{equation}
Suppressing the uncertainty $\sigma_V$ below the $\epsilon$-fraction of the signal, $1-V^{\rm eff}$, requires
\begin{equation}\label{epsilon control of shot noise}
    \frac{\sigma_V}{1-V^{\rm eff}}\leq\epsilon \quad \Longrightarrow \quad N_{\rm det}\geq \frac{1}{\epsilon^2}\frac{1+V^{\rm eff}}{1-V^{\rm eff}}.
\end{equation}
The count rate at a single monitored output, averaged over equal acquisition times at the fringe maximum and minimum, is
\begin{equation}
    \overline{\dot N}_{\rm det}=\frac{1}{2}\beta R_{\rm avg}^{\rm eff}\dot N_{\rm in}.
\end{equation}
Assuming that \(N_{\rm det}=\overline{\dot N}_{\rm det}T\), where \(T\) is the integration time, Eq.~\eqref{epsilon control of shot noise} yields the following requirement
\begin{equation}
    T\geq\frac{2}{\epsilon^2\alpha\Gamma_{\rm P}\beta R_{\rm avg}^{\rm eff}}\frac{1+V^{\rm eff}}{1-V^{\rm eff}}.
\end{equation}

For a concrete numerical estimate, we consider the reference signal used in Table~\ref{tab:noise-requirements}, corresponding to \(V^{\rm eff}\simeq0.232\) and \(R_{\rm avg}^{\rm eff}\simeq0.656\).
For \(C=1.1\), the total Purcell-enhanced population-decay rate is \(\Gamma_{\rm P}\simeq2.71\times10^{-2}\,{\rm s}^{-1}\).
We assume the incident flux parameter \(\alpha=0.2\) and the propagation and detection efficiency \(\beta=0.5\).
Suppressing $\sigma_V$ below one-third of the predicted visibility loss ($\epsilon=1/3$) requires integration time $T\gtrsim4.50\,{\rm h}$.
This estimate includes photon shot noise only; locating and tracking the fringe extrema, acquiring an independent reference measurement, and accounting for background counts and technical drifts would increase the required experimental time.

\subsection*{Use of AI tools}
ChatGPT (OpenAI) assisted with language, presentation, and discussion of the theoretical analysis.
Claude (Anthropic) suggested an analytical solution to the finite-bandwidth model in terms of Faddeeva functions.
The authors reviewed all suggestions and take full responsibility for the final manuscript.

\section*{Acknowledgments}
J.P. and M.Z. thank Navdeep Arya and Patryk Michalski for helpful discussions.
J.P. and M.Z. acknowledge the Knut and Alice Wallenberg Foundation through a Wallenberg Academy Fellowship No.\ 2021.0119.
P.T.G. was supported by the project CZ.02.01.01/00/22\_010/0013054 (C-MONS) within Programme JAC MSCA Fellowships at Palacký University Olomouc IV.
S.W. acknowledges support from the Center for Integrated Quantum Science and Technology (IQST) and financial support from the German Research Foundation through Emmy Noether Grant No. WE 7554/1-1, and the Carl-Zeiss-Stiftung Center for Quantum Photonics (QPhoton).
N.B. acknowledges support from ASI and CNR under the Joint Project “Laboratori congiunti ASI-CNR nel settore delle Quantum Technologies (QASINO)” (Accordo Attuativo n. 2023-47-HH.0).
\section*{Author contributions}
J.P. conceived the original idea and initiated the theoretical analysis.
All authors contributed to developing the theoretical framework, interpreting the results, and writing and revising the manuscript.
M.Z. supervised the project.
%\section*{Data availability}
\bibliography{GravInter}
\iftoggle{arXiv}{
	\clearpage
	\title{Supplementary Material: Spectrally enhanced gravitational redshift signals in photonic interferometry}
    \maketitle
}{}
\label{supplement}
\setcounter{figure}{0}
\setcounter{equation}{0}
\renewcommand{\theequation}{S\arabic{equation}}
\renewcommand{\theHequation}{S\arabic{equation}}
\renewcommand{\thefigure}{S\arabic{figure}}

\section{Analytical evaluation of the finite-bandwidth response}
We derive here the closed-form expressions for the frequency-averaged reflection probabilities and interference amplitude for a finite-bandwidth Gaussian probe.
We consider two identical atom–cavity scatterers with ideal (lossless) single-sided cavities and use the effective reflection amplitude from the main text, neglecting cavity detuning over the relevant spectral range:
\begin{equation}
    r(\Delta)=-1+\frac{4C\gamma}{\Gamma_{\rm P}/2+\mathrm{i}\Delta}.
\end{equation}
Let \(\Omega=\omega-\bar{\omega}\) denote the angular-frequency offset from the central probe frequency.
The normalized spectral probability distribution is
\begin{equation}
    p(\Omega)\equiv|f(\omega)|^2=\frac{1}{\sqrt{2\pi}\sigma}\exp\left(-\frac{\Omega^2}{2\sigma^2}\right)
\end{equation}
We denote the central probe detunings at the lower and upper scatterers by \(\Delta_{\rm l}\) and \(\Delta_{\rm u}\), respectively, so that a spectral component with offset \(\Omega\) probes the two scatterers at detunings \(\Delta_{\rm l}+\Omega\) and \(\Delta_{\rm u}+\Omega\).
The quantities to be evaluated are therefore
\begin{equation}
    \begin{split}
        &R_j^{\rm eff}=\int_{-\infty}^{\infty}\dd\Omega p(\Omega)\left|r(\Delta_j+\Omega)\right|^2,
        \qquad j\in\{{\rm l},{\rm u}\},\\
        &\mathcal{A}^{\rm eff}=\int_{-\infty}^{\infty}\dd\Omega p(\Omega) r^\ast(\Delta_{\rm l}+\Omega) r(\Delta_{\rm u}+\Omega).
    \end{split} 
\end{equation}

As we shall see, both can be expressed in terms of the dimensionless auxiliary function
\begin{equation}
    \mathcal{G}(\Delta)=\int_{-\infty}^{\infty}\dd\Omega p(\Omega)\frac{\gamma}{\Gamma_{\rm P}/2+\mathrm{i}(\Delta+\Omega)}.
\end{equation}
To evaluate this integral, we substitute
\begin{equation}
    t=\frac{\Omega}{\sqrt{2}\sigma},
    \qquad
    z(\Delta)=\frac{\mathrm{i}\Gamma_{\rm P}/2-\Delta}{\sqrt{2}\sigma},
\end{equation}
which gives
\begin{equation}
    \begin{split}
        \mathcal G(\Delta)&=\frac{\mathrm{i}\gamma}{\sqrt{2\pi}\sigma}\int_{-\infty}^{\infty}\dd t\frac{e^{-t^2}}{z(\Delta)-t}\\&=\sqrt{\frac{\pi}{2}}\frac{\gamma}{\sigma}w\left(z(\Delta)\right).
    \end{split}
\end{equation}
Here, in the last step, we identified the Faddeeva function~\cite{DLMF} \(w(z)=\ee^{-z^2}\operatorname{erfc}(-\ii z)\), which, for \(\Im(z)>0\), has the integral representation
\begin{equation}
    w(z)=\frac{\mathrm{i}}{\pi}\int_{-\infty}^\infty\dd t\frac{\ee^{-t^2}}{z-t}.
\end{equation}

Let us now show that \(R_j^{\rm eff}\) and \(\mathcal{A}^{\rm eff}\) can be expressed in terms of \(\mathcal{G}(\Delta)\).
For the first one, we note that the reflection probability \(R(\Delta)\equiv|r(\Delta)|^2\) from the main text can be rewritten as follows:
\begin{equation}
    |r(\Delta)|^2=1-\frac{8C}{1+2C}\Re\left[\frac{\gamma}{\Gamma_{\rm P}/2+\mathrm{i}\Delta}\right].
\end{equation}
Using \(\int_{-\infty}^\infty\dd\Omega p(\Omega)=1\) and
\begin{equation}
    \int_{-\infty}^\infty\dd\Omega p(\Omega)\Re\left[\frac{\gamma}{\Gamma_{\rm P}/2+\mathrm{i}(\Delta+\Omega)}\right]=\Re\left[\mathcal{G}(\Delta)\right],
\end{equation}
we find
\begin{equation}\label{supp: rj}
    R_j^{\rm eff}=1-\frac{8C}{1+2C}\Re\left[\mathcal{G}(\Delta_j)\right],
\end{equation}
and consequently
\begin{equation}\label{supp: ravg}
    R_{\rm avg}^{\rm eff}=1-\frac{4C}{1+2C}\Re\left[\mathcal{G}(\Delta_{\rm l})+\mathcal{G}(\Delta_{\rm u})\right].
\end{equation}
For the interference amplitude \(\mathcal{A}^{\rm eff}\), we use
\begin{equation}
    \begin{split}
        r^\ast(\Delta_1)r(\Delta_2)=&1-4C\frac{2\gamma+\mathrm{i}(\Delta_2-\Delta_1)}{\Gamma_{\rm P}+\mathrm{i}(\Delta_2-\Delta_1)}\times\\
        \times&\left(\frac{\gamma}{\Gamma_{\rm P}/2-\mathrm{i}\Delta_1}+\frac{\gamma}{\Gamma_{\rm P}/2+\mathrm{i}\Delta_2}\right),
    \end{split}
\end{equation}
which, combined with \(\Delta_{\rm u}-\Delta_{\rm l}=\delta\omega_g\), yields
\begin{equation}\label{supp: aeff}
    \mathcal{A}^{\rm eff}=1-4C\left[\mathcal{G}(\Delta_{\rm u})+\mathcal{G}^\ast(\Delta_{\rm l})\right]\frac{2\gamma+\mathrm{i}\delta\omega_g}{\Gamma_{\rm P}+\mathrm{i}\delta\omega_g}.
\end{equation}
In the limit of probe bandwidth much smaller than the Purcell-broadened linewidth, \(\sigma/\Gamma_{\rm P}\to0\), we have
\begin{equation}
    \mathcal{G}(\Delta)\to\frac{\gamma}{\Gamma_{\rm P}/2+\mathrm{i}\Delta},
\end{equation}
and so we recover the monochromatic expressions
\begin{equation}
    R_j^{\rm eff}\to|r(\Delta_j)|^2,\qquad \mathcal{A}^{\rm eff}\to r^\ast(\Delta_{\rm l})r(\Delta_{\rm u}).
\end{equation}

Using Eqs.~\eqref{supp: ravg} and~\eqref{supp: aeff}, we numerically maximize the visibility loss \(1-V^{\rm eff}\) and phase magnitude \(|\Phi_g^{\rm eff}|\), where
\begin{equation}
1-V^{\rm eff}=1-\frac{|\mathcal{A}^{\rm eff}|}{R_{\rm avg}^{\rm eff}},
\qquad
\Phi_g^{\rm eff}=\arg\mathcal{A}^{\rm eff}.
\end{equation}
As in the main text, we consider the \({}^{87}\mathrm{Sr}\) clock transition, with \(\gamma=2\pi\times0.675\,\mathrm{mHz}\), and a Gaussian probe with bandwidth \(\sigma=6.29\gamma\), corresponding to \(\Delta\nu_{\rm FWHM}=10\,\mathrm{mHz}\).
Scanning the height separation, cooperativity, and lower-arm detuning over \(h\in[0,10]\,\mathrm{cm}\), \(C\in[0,5]\), and \(\Delta_{\rm l}/\gamma\in[-10,10]\), respectively, we find both maxima at \(h=10\,\mathrm{cm}\) and symmetric detuning, \(\Delta_{\rm l}=-\delta\omega_g/2\).
They occur at different cooperativities: the visibility loss reaches \(1-V^{\rm eff}\simeq0.77\) at \(C\simeq1.10\), whereas the phase magnitude reaches \(|\Phi_g^{\rm eff}|\simeq0.50\pi\) at \(C\simeq1.86\).
Fixing \(h=10\,\mathrm{cm}\) and extending the search to \(C\in[0,100]\) and \(\Delta_{\rm l}/\gamma\in[-10^5,10^5]\) yields the same optima; no larger values were found within this extended search domain.

\section*{Mach--Zehnder interferometer}
\begin{figure*}[t]
    \centering
    \includegraphics[width=\textwidth]{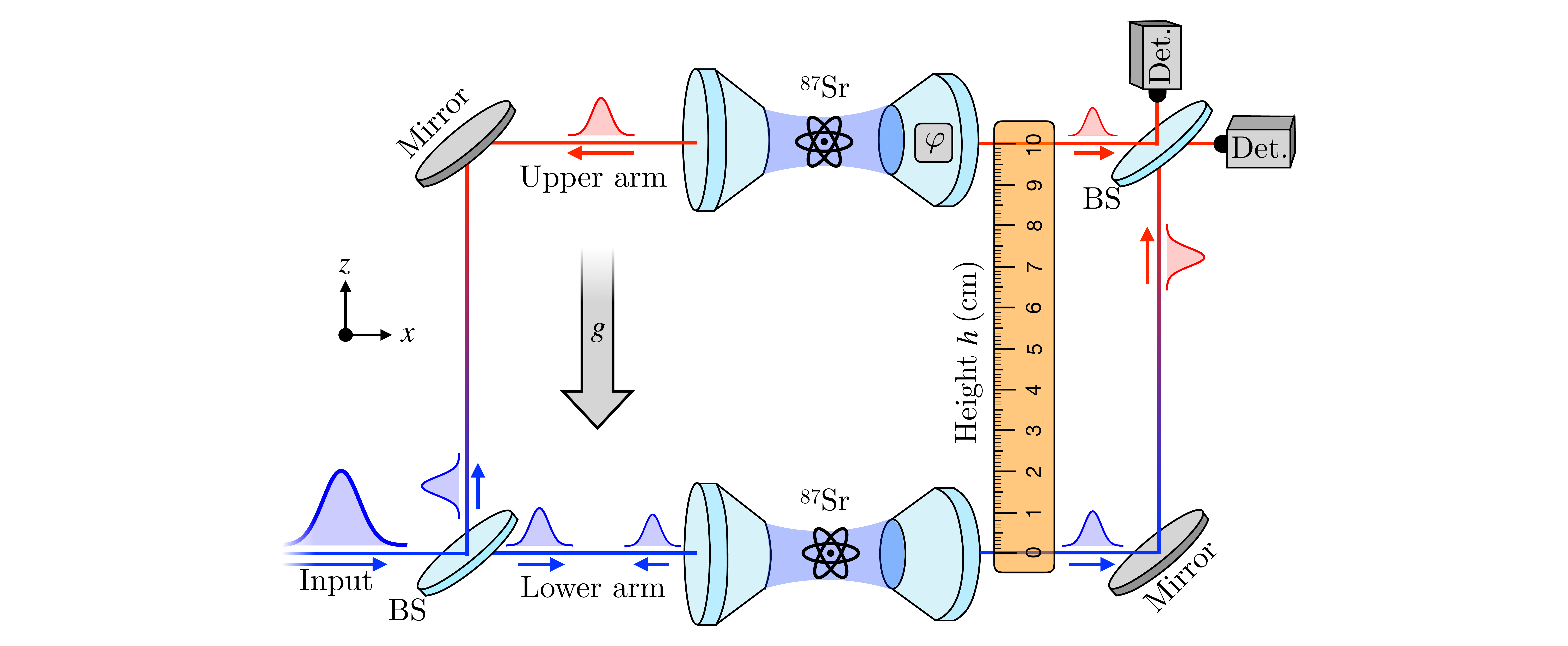}
    \caption{\justifying Mach-Zehnder interferometer with optical cavities inserted in each arm, and an auxiliary phase shift ($\varphi$) introduced in the upper arm. The gravitational field causes a redshift of the photons traveling upwards, introducing a difference between transmission amplitudes in both arms, and, consequently, affecting the total phase shift between the arms and the visibility of interference.}
    \label{fig:Fig_S1}
\end{figure*}
%\vspace{-4.5em}
\begin{figure*}[t]
    \centering
    \includegraphics[width=\textwidth]{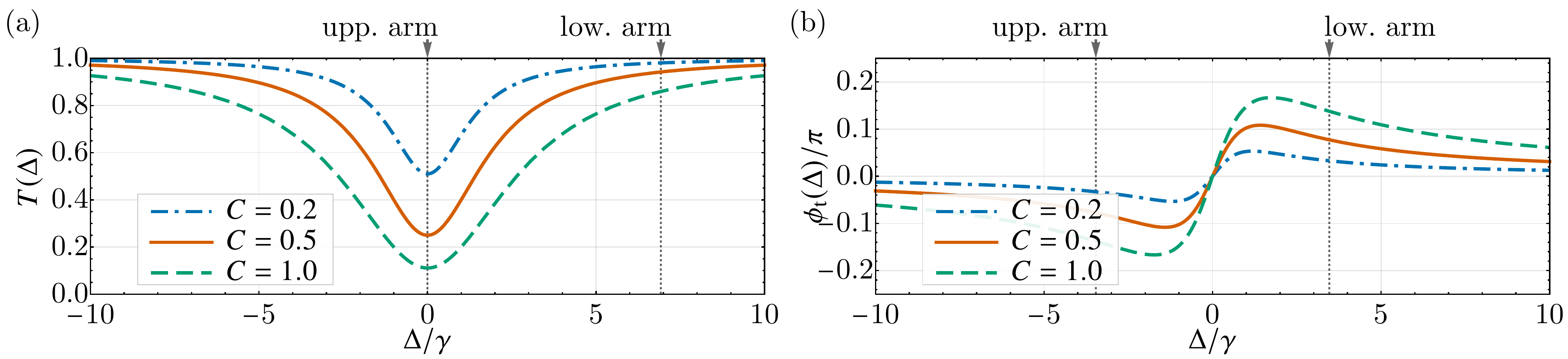}
    \caption{\justifying Transmission probability (a) and transmission phase (b) as functions of the photon detuning $\Delta$ for a lossless symmetric cavity-QED scatterer. Both transmission probability and phase are plotted for three different values of the atom--cavity cooperativity $C\in\{0.2,0.5,1\}$. Dotted vertical lines mark the frequency of the photon in the lower and upper arms for the frequency shift between the arms $\delta\omega_g/\gamma\simeq-6.93$, which, for the ${}^{87}\mathrm{Sr}$ clock transition, corresponds to the height difference $h=10\,{\rm cm}$. In (a), the photon is assumed to be in resonance in the upper arm --- then it is out of resonance in the lower arm. In (b), we assume a symmetric detuning $\Delta_{\rm u}=-\Delta_{\rm l}=\delta\omega_g/2$.}
    \label{fig: transmission phase and probability}
\end{figure*}
In analogy with the Michelson interferometer discussed in the main text, here we consider a Mach--Zehnder geometry with identical cavity-QED scatterers inserted in the two arms, as shown in Fig.~\ref{fig:Fig_S1}.
For each arm, we denote the complex transmission amplitude by
\begin{equation}
t_{\rm l,u}=\sqrt{T_{\rm l,u}}\exp\left(\ii\phi_{\rm l,u}\right),
\end{equation}
where $\rm l$ and $\rm u$ label the lower and upper arms, respectively.
We also include an externally controlled phase $\varphi$ in the upper arm.
Using the same beam-splitter convention as in the main text, the unnormalised transmitted component of the photon state after recombination at the second beam splitter is
\begin{equation}
\ket{\psi_f}=\frac{1}{2}\left[\left(t_{\rm l}-\ee^{\ii\varphi}t_{\rm u}\right)\ket{0}+\ii\left(t_{\rm l}+\ee^{\ii\varphi}t_{\rm u}\right)\ket{1}\right],
\end{equation}
where $\ket{0}$ and $\ket{1}$ denote the two output ports.
The state is unnormalised because only the transmitted components of the two cavity-QED scatterers contribute to this output.
Its norm,
\begin{equation}
\braket{\psi_f}{\psi_f}=\frac{T_{\rm l}+T_{\rm u}}{2}\equiv T_{\rm avg},
\end{equation}
therefore quantifies the total transmitted signal.
The probability of detecting the photon in output port $i\in\{0,1\}$ is
\begin{equation}
    P_i(\varphi)=\frac{T_{\rm avg}}{2}\left[1-(-1)^i\frac{\sqrt{T_{\rm l}T_{\rm u}}}{T_{\rm avg}}\cos\left(\varphi+\phi_{\rm u}-\phi_{\rm l}\right)\right].
\end{equation}
Unlike in the Michelson geometry, both output ports of the Mach-Zehnder interferometer are directly accessible.
We identify
\begin{equation}
V=\frac{\sqrt{T_{\rm l}T_{\rm u}}}{T_{\rm avg}},\qquad\Phi_g=\phi_{\rm u}-\phi_{\rm l},
\end{equation}
as the interference visibility and the gravity-induced phase difference, respectively.
The transmission amplitude in the weak-excitation regime is given by~\cite{Reiserer2015}
\begin{equation}\label{transmission amplitude}
    t(\omega)=\frac{2\sqrt{\kappa_1\kappa_2}(\mathrm{i}\Delta_{\rm a}+\gamma)}{(\mathrm{i}\Delta_{\rm c}+\kappa)(\mathrm{i}\Delta_{\rm a}+\gamma)+g_0^2},
\end{equation}
where $\kappa_1$ and $\kappa_2$ are the decay rates through the left and right mirrors, respectively, $\kappa=\kappa_1+\kappa_2+\kappa_{\rm loss}$ is the total cavity decay rate, and other parameters are defined as in the main text.
We assume that the bare cavity response is flat over the relevant frequency range, and that the atom is resonant with the cavity, $\Delta_{\rm a}=\Delta_{\rm c}\equiv\Delta$.
Furthermore, we focus on photons with frequencies around this resonance, such that $|\Delta|\sim\gamma\ll\kappa$, meaning that we can set $\Delta/\kappa\simeq0$.
Specifying the considerations to a lossless symmetric cavity, $\kappa_{\rm loss}=0$ and $\kappa_1=\kappa_2=\kappa/2$, we can rewrite the transmission amplitude in the simplified form
\begin{equation}
    t(\Delta)=\frac{1+\ii\Delta/\gamma}{1+\ii\Delta/\gamma+2C}.
\end{equation}
This leads to the following expressions for the transmission probability and phase:
\begin{equation}\label{cQED transmission probability and phase}
    \begin{split}
        &T(\Delta)=\frac{1+(\Delta/\gamma)^2}{(1+2C)^2+(\Delta/\gamma)^2},\\
        &\phi_t(\Delta)=\arctan\left(\frac{2C\Delta/\gamma}{1+2C+(\Delta/\gamma)^2}\right).
    \end{split}
\end{equation}
Both functions are plotted in Fig.~\ref{fig: transmission phase and probability} for different values of cooperativity $C$.

The transmission probability in Fig.~\ref{fig: transmission phase and probability} (a) resembles the reflection probability of a single-sided cavity-QED scatterer discussed in the main text.
It shows a dip centered on resonance, $\Delta=0$, with a full width at half-maximum depth $\Gamma_{\rm P}=2(1+2C)\gamma$.
However, the depth of this dip, $1-T(0)=4C(1+C)/(1+2C)^2$, is a monotonically growing function of $C$, and tends to unity in the limit of $C\to\infty$; in contrast, the same, maximal depth for the reflection spectrum is achieved at a finite cooperativity $C=0.5$.
The acquired phase, shown in Fig.~\ref{fig: transmission phase and probability} (b), has two extrema: a minimum at $\Delta=-\sqrt{1+2C}\gamma$ and a maximum at $\Delta=\sqrt{1+2C}\gamma$, both with magnitude $|\phi_t|=\arctan(C/\sqrt{1+2C})$.
For large detunings, $|\Delta|\gg\sqrt{1+2C}\gamma$, the acquired phase tends to 0.
Consider the concrete example of the ${}^{87}\mathrm{Sr}$ ultranarrow clock transition, ${}^3P_0\leftrightarrow{}^1S_0$, and height difference $h=10\,{\rm cm}$, corresponding to the gravitational redshift $\delta\omega_g/\gamma\simeq-6.93$.
The two relevant operational points are depicted in Fig.~\ref{fig: transmission phase and probability}: upper-arm resonance, $\Delta_{\rm u}=0$ and $\Delta_{\rm l}=-\delta\omega_g$, shown in panel (a), and symmetric detuning, $\Delta_{\rm u}=-\Delta_{\rm l}=\delta\omega_g/2$, shown in panel (b).
Figure~\ref{fig: transmission phase and probability} shows that a significant difference in transmission probability and phase can be achieved already for moderate values of cooperativity.
For example, for $C=0.5$ and upper-arm resonance, the transmission probability in the upper arm would be $T_{\rm u}=0.25$, while in the lower arm we would have $T_{\rm l}\simeq0.94$; this would lead to the gravity-induced visibility loss $1-V\simeq0.19$, with the total transmitted signal $T_{\rm avg}\simeq0.60$.
On the other hand, for the symmetric detuning and the same cooperativity, the phase difference between the two arms would be $|\Phi_g|\simeq0.15\pi$, with the total transmitted signal $T_{\rm avg}\simeq0.81$.
Comparing with the results from the main text, we conclude that the same value of cooperativity leads to a larger visibility loss and phase difference for the reflection-based scheme, at the price of a smaller output signal.

\section*{Atoms coupled to a waveguide}
As noted in the main text, atoms coupled directly to a waveguide constitute a promising alternative to cavity-QED scatterers~\cite{Sheremet2023}.
In the weak excitation regime, the reflection and transmission amplitudes for a single atom coupled symmetrically to a waveguide are given by~\cite{Sheremet2023}
\begin{equation}\label{atom reflection and transmission}
    \begin{split}
        &r(\Delta)=-\frac{\ii\gamma_{\rm 1D}}{\Delta+\ii(\gamma+\gamma_{\rm 1D})},\\
        &t(\Delta)=\frac{\Delta+\ii\gamma}{\Delta+\ii(\gamma+\gamma_{\rm 1D})},
    \end{split}
\end{equation}
where $\Delta=\omega-\omega_{\rm a}$ is the photon-atom detuning, and $\gamma_{\rm 1D}$ and $\gamma$ denote the contributions to the atomic polarization decay rate from guided and nonguided modes, respectively.
Reorganizing the imaginary factors, rescaling the numerators and denominators by $\gamma$, and introducing atom-waveguide cooperativity $\tilde{C}=\gamma_{\rm 1D}/\gamma$, we can rewrite Eq.~\eqref{atom reflection and transmission} as follows:
\begin{equation}
    \begin{split}
        &r(\Delta)=-\frac{\tilde{C}}{1-\ii\Delta/\gamma+\tilde{C}},\\
        &t(\Delta)=\frac{1-\ii\Delta/\gamma}{1-\ii\Delta/\gamma+\tilde{C}}.
    \end{split}
\end{equation}
These lead to the following transmission and reflection probabilities:
\begin{equation}
    \begin{split}
        &R(\Delta)=\frac{\tilde{C}^2}{\left(1+\tilde{C}\right)^2+\left(\Delta/\gamma\right)^2},\\
        &T(\Delta)=\frac{1+\left(\Delta/\gamma\right)^2}{\left(1+\tilde{C}\right)^2+\left(\Delta/\gamma\right)^2},
    \end{split}
\end{equation}
and the corresponding phases
\begin{equation}
    \begin{split}
        &\phi_r(\Delta)=\pi+\arctan\left(\frac{\Delta/\gamma}{1+\tilde{C}}\right),\\
        &\phi_t(\Delta)=-\arctan\left(\frac{\tilde{C}\Delta/\gamma}{1+\tilde{C}+\left(\Delta/\gamma\right)^2}\right).
    \end{split}
\end{equation}
Notably, up to a replacement $\tilde{C}\to2C$, the transmission probability is the same as for the symmetric cavity-QED system discussed in the previous section, while the associated phase differs only by a sign (see Eq.~\eqref{cQED transmission probability and phase}; the sign difference of the phase comes from different conventions in Refs~\cite{Reiserer2015} and~\cite{Sheremet2023}).

The main conceptual advantage of the waveguide-QED implementation is that the cavity resonance is removed as an additional spectral element.
In the simple bidirectional waveguide model considered here, the response is set directly by the atomic resonance and by the coupling ratio \(\tilde{C}\).
This eliminates cavity-detuning noise and may be attractive for integrated photonic implementations.
On the other hand, waveguide platforms replace cavity-related noise by a different set of implementation-specific constraints associated with atom trapping and coupling near nanophotonic structures, including position-dependent coupling, polarization-dependent guided modes, fabrication disorder, surface adsorption, and Casimir--Polder surface potentials~\cite{Chang2018}.
Finally, while the millihertz-linewidth \({}^{87}\mathrm{Sr}\) clock transition has already been coupled to high-finesse optical cavities~\cite{Norcia2016, Norcia2018, Muniz2021}, an analogous waveguide-QED implementation on this transition has not yet been demonstrated.
Thus, atoms coupled to waveguides provide a natural and potentially compact extension of the present idea, but cavity-QED with strontium currently offers the more experimentally grounded route to an ultranarrow scattering response.

\end{document}